\documentclass{IEEEtran}
\usepackage{cite}
\usepackage{amsmath,amssymb,amsfonts}
\usepackage{algorithmic}
\usepackage{graphicx}
\usepackage{textcomp}
\usepackage[justification=centering]{caption}
\usepackage{float}
\usepackage{booktabs}
\usepackage{multirow}
\usepackage{multicol}
\usepackage{setspace}
\usepackage{threeparttable}
\usepackage{xcolor}
\usepackage{subfigure}
\usepackage[ruled,vlined]{algorithm2e}

\newcommand{\algrule}[1][.5pt]{\par\vskip.5\baselineskip\hrule height #1\par\vskip.5\baselineskip}

\def \etal {\emph{et al. }}
\def \eq {Eq.}
\def \f {Fig.}

\def\BibTeX{{\rm B\kern-.05em{\sc i\kern-.025em b}\kern-.08em
    T\kern-.1667em\lower.7ex\hbox{E}\kern-.125emX}}
\begin{document}
\title{Spectrum Congruency of Multiscale Local Patches for Edge Detection}
\author{\IEEEauthorblockN{Fang Yang$^1$, Xin Su$^2$*, Li Chai$^1$}

\IEEEauthorblockA{\textit{1 Engineering Research Center of Metallurgical Automation and Measurement Technology,} \textit{Wuhan University of Science and Technology}
430081, Wuhan, China \\}
\textit{2 School of Remote Sensing and Information Engineering}
\textit{Wuhan University}, 430079, Wuhan, China \\
}
\maketitle

\begin{abstract}

This paper proposes a novel feature called spectrum congruency for describing edges in images.
The spectrum congruency is a generalization of the phase congruency, which depicts how much each Fourier components of the image are congruent in phase.
Instead of using fixed bases in phase congruency, the spectrum congruency measures the congruency of the energy distribution of multiscale patches in a data-driven transform domain, which is more adaptable to the input images.
Multiscale image patches are used to acquire different frequency components for modeling the local energy and amplitude.
The spectrum congruency coincides nicely with human visions of perceiving features and provides a more reliable way of detecting edges.
Unlike most existing differential-based multiscale edge detectors which simply combine the multiscale information, our method focuses on exploiting the correlation of the multiscale patches based on their local energy.
We test our proposed method on synthetic and real images, and the results demonstrate that our approach is practical and highly robust to noise.
\end{abstract}


\section{Introduction}
\label{sec:introduction}

Edge detection is one of the most fundamental and essential tasks in the computer vision field.
Edges play essential roles and act as the basis of a number of tasks such as image segmentation and object recognition \cite{koschan2005detection,amfm_pami2011,chen2012,2013Edge,2015A,2016Noise}.
In the views of image processing, an edge is a place where there is a sharp change or discontinuity of intensity or brightness.
To find the discontinuity, most edge detection methods focus on the maximum of the absolute values of the first-order derivative or the zero crossings of the second-order derivative.
Besides, some methods based on the signal decomposition or wavelets are also favored in edge detection \cite{lopezmolina2013multiscale,Imamoglu2013}.

The last decades have witnessed a huge growth in the development of edge detection methods.
The existing edge detectors can be roughly divided into two classes: single scale-based and multiple scale-based methods \cite{2019Multiscale}.
Chronologically, early single-scale edge detection approaches aim at identifying points where the image brightness changes sharply, such as Robert operator \cite{roberts1963machine}, Sobel operator \cite{sobel1970camera}, Prewitt operator \cite{prewitt1970object} \emph{etc}.
These detectors use simple gradient operators to detect edges and are very sensitive to noise, in the meantime, they can not provide precise locations of edges.
Then Marr and Hildreth proposed to search for the the zero crossings of the second order derivative of an image, which helps to improve the performance \cite{marr1980theory}.
However, the Marr-Hildreth algorithm does not behave well at places where the intensity varies, and it generates localization errors.
The Canny detector is the most widely used edge detector since it was proposed \cite{Canny86}.
Canny detector aims at finding an optimal edge detection algorithm, which means that the algorithm 1) could label as many real edges as possible in images; 2) could localize the edges accurately on the edge; 3) should mark no more than once for an edge and should not mark the possible noise as the edges.
Canny detector preprocess the image with a Gaussian kernel, which acts as a scale selection.
With a small scale Gaussian kernel, the Canny detector can provide details of intensity variation information, but it is prone to noise.
A large scale Gaussian kernel makes the Canny detector robust to noise, but the precision of location is degraded more or less \cite{deriche1987using,ding2001on}.
The single-scale edge detectors are actually defined to detect step edges.
However, in natural images, one can generally find spatially scaled edges of which the intensity discontinuities vary over different widths \cite{elder1998local}.
Hence, single-scale edge detectors do not conform to actual situations and are not sufficient for edge detection.
Then multiscale techniques are employed to describe and synthesize the varieties of edges \cite{Torre1986On,Bezdek2002A}.

Multiscale approaches can be implemented either in the spatial domain or in the transform domain.
In the spatial domain, the edge strength maps are usually obtained by differential-based methods, as done in the single scale.
For example, Bao \etal proposed a scale multiplication function to fuse the responses of Canny detector at different scales \cite{Bao2005Canny}.
Shui \etal proposed an edge detection method based on the multiplication of isotropic and anisotropic Gaussian directional derivatives \cite{2012Noise}.
Wang \etal detected the edges by using the first-order derivative of the anisotropic Gaussian kernel, which improves the robustness to noise for small scale kernels \cite{2019Multiscale}.
In \cite{lopezmolina2013multiscale}, the authors proposed a multiscale edge detection in the Gaussian scale space based on Gaussian smoothing and put forward a coarse-to-fine tracking algorithm to combine the results at each scale.
However, differential-based edge detection methods are always sensitive to variations in image illumination and it is hard to determine the threshold values.
Besides, only a very few works aim at exploring the fusion of multiscale information \cite{lopezmolina2013multiscale}.
Except for the scale multiplication, morphological dilation and multiplication was also proposed to fuse the multiscale information \cite{papari2007a}.
These methods simply combine the multiscale formation but do not provide an inner and deeper explanation.
In our paper, to better exploit the correlation between multiscale information, we adopt to fuse the multiscale information in the transform domain according to the local energy model.

In \cite{morrone1986mach, morrone1987feature}, the authors found that biologically or physically, the edges in images could be defined as places where the Fourier components are maximally congruent in phase.
This phenomenon is called the phase congruency.
It is proved in \cite{concetta1988feature} that the phase congruency is consistent with the human visual system, \emph{i.e.}, it explains a number of psychophysical effects that how human perceive the features in images \cite{kovesi1999image,kovesi2000phase}.
Therefore, it is invariant to image brightness and illumination.
Basically, the phase congruency is computed in the following three stages: 1) use several pairs of quadrature filters with different scales to obtain the low-, mid- and high-frequency filter templates; 2) convolve the templates with the input image, thus bringing the information at each frequency band; 3) compute the phase congruency according to the pixel local energy, amplitude, the estimated noise energy and a weighting function.

Phase congruency can be comprehended as salient features that have similar values of the local phase when observed at different scales \cite{Obara2012Coherence}.
It can be also regarded as the edge strength map.
As a low-level image feature, the phase congruency is favored and widely applied in many kinds of image vision tasks because of its invariance to illumination and consistency to the human visual system, such as image quality assessment \cite{zhang2011fsim}, multi-modal image analysis \cite{Bhatnagar2013, Li2020}, pattern recognition and object detection \cite{Gundimada09,verikas2012phase,shojaeilangari2014novel,Mouats15}, image registration \cite{ye2017robust,fan2018sar} and so on.
However, existing measurements of phase congruency are sensitive to noise, even though they take the noise compensation into consideration.
Moreover, they generate glitch artifacts when the input image is noisy because they integrate the response value from multiple orientations, which may lead to spurious edges \cite{kovesi2000phase,kovesi2003phase}.
Besides, the phase congruency model extracts information of different frequency bands by convolving the image with Log-Gabor filters of different scales, however, the Log-Gabor filter is characterized by Gaussian kernels, which reduces some high-frequency components.
Therefore, a weighting function is needed to devalue phase congruency at locations where the spread of filter responses is narrow \cite{kovesi2000phase}, which may reduce the precision of feature location.

In this paper, we are motivated to find a proper data-driven transform domain to represent the input image and compute the local energy.
The new measurement is called spectrum congruency and it is a generalization of the phase congruency.
With the data-driven bases, we do not need to integrate the filter response value from multiple orientations, thus avoiding the glitch artifact.
In addition, the spectrum congruency extracts different frequency information by using multiscale patches and can retain all frequency information, hence, the weighting function is unnecessary.
The spectrum congruency is more adaptable to the input image, which makes the detected features more reliable.
To the best of our knowledge, there are no results in the literature regarding applying the data-driven method to measure the local energy and the phase congruency.

In recent years, with the rapid development of big data, the learning-based \cite{amfm_pami2011,lim2013sketch,Dollar13,zhang2016semicontour} edge detection methods, especially  deep learning-based approaches \cite{ganin2014n,bertasius2015deepedge,shen2015deepcontour,xie2015holistically,he2019bidirectional} have achieved great success.
Despite the superior performance of the learning-based methods, it is generally acknowledged that such techniques require a large amount of training data.
Moreover, they suffer from a lack of interpretability and their computation is complicated, time-consuming.
Therefore, it is worth developing the model-based edge detection method to quickly obtain accurate results without supervising information and training data.

The remainder of this paper is organized as follows.
Section \ref{sec:localenergymodel} introduces the local energy model and how it is applied to compute the phase congruency feature.
Section \ref{sec:proposedmethod} presents our proposed spectrum congruency according to data-driven local energy model in detail.
Section \ref{sec:experiment} analyzes the effect of patchsizes and demonstrates the effectiveness of our method on real image data.
Section \ref{sec:conclusion} concludes the paper.

\section{Local Energy Model and Phase Congruency}
\label{sec:localenergymodel}
The original way to calculate phase congruency of a signal is to compute the phase difference of each Fourier components, which is not efficient.
It is indicated in \cite{venkatesh1990classification} that points of maximum phase congruency are located in the peaks of local energy.
Therefore, to simplify the calculation, Venkatesh and Owens proposed to measure phase congruency according to the local energy $E(x)$:
\begin{equation}
E(x)=\sqrt{F^2(x)+H^2(x)},
\end{equation}
where $F(x)$ is the signal $f(x)$ without its DC component, and $H(x)$ is the Hilbert transform of $F(x)$.
The phase congruency of the signal is equivalent to its local energy scaled by the reciprocal of the sum of Fourier amplitude \cite{venkatesh1990classification}:
\begin{equation}
\label{eq:energy}
PC(x)=\frac{E(x)}{\sum_n A_n(x)+\epsilon},
\end{equation}
where $A_n(x)$ represents the amplitude of the $n_{th}$ Fourier component and $\epsilon$ is a small positive value to keep the denominator from being zero.

However, the Fourier transform is insufficient to determine local frequency information \cite{kovesi1999image}.
This is because that the Fourier transform does not take the spread of frequencies into account, which makes it invalid for some special signals, \emph{e.g.,} signals with only one frequency component.
Another important reason is that windowing is a necessary operation to control the scale, which makes the method complicated.
Afterwards, wavelet transform are used to calculate the phase congruency respectively \cite{kovesi1999image}.
The log-Gabor filter is applied because it can be constructed with arbitrary bandwidth and can obtain local frequency information appropriately.
The local energy based on the log-Gabor filter is computed as:
\begin{equation}\label{EnergyLogGabor}
E(x)=\sqrt{(\sum_ne_n(x))^2+(\sum_no_n(x))^2},
\end{equation}
where $e_n(x)$ and $o_n(x)$ are the even and odd filter response at $n^{th}$ scale.
The corresponding amplitude of the signal at each frequency scale is :
\begin{equation}\label{AmplitudeLogGabor}
A_n(x) = \sqrt{(e_n(x)^2+o_n(x)^2)}
\end{equation}

A 2D Gabor filter is actually a Gaussian kernel function modulated by a sinusoidal plane wave.
Hence, when applying the log-Gabor filter to the input image, some high-frequency information is removed, which will lead to spurious responses.
Thus, a weighting function $W(\cdot)$ was constructed by applying a sigmoid function to the filter response value and multiplied to the local energy.
Additionally, the noise compensation was taken into account by subtracting the estimated noise energy $T$ from the local energy.
The phase congruency model that is computed by log-Gabor filters then becomes:
\begin{equation}\label{eq:PC2}
    PC(x)=\frac{W(x)\lfloor E(x) - T \rfloor}{\sum_n A_n(x)+\epsilon},
\end{equation}
where $\lfloor \cdot \rfloor$ denotes that the enclosed value is itself if it is positive, and otherwise zero.
It is postulated that the distribution of the magnitude of energy vector has a Rayleigh distribution, hence, $T$ is the estimated noise energy according to the mean value and variance of the Rayleigh distribution at the smallest scale.

The classical phase congruency model computes the energy and amplitude in the transform domain with fixed bases, such as the Fourier transform \cite{venkatesh1990classification}, the log-Gabor transform \cite{kovesi2000phase}, the monogenic signal \cite{felsberg2001monogenic}, which results in spurious edges and glitch artifacts in noisy images.
Therefore, we would like to develop a data-driven method that is adaptable to the input images to compute the edge feature map.
\section{Spectrum Congruency and Edge Detection}
\label{sec:proposedmethod}

\subsection{Spectrum Congruency via Multiscale Local Energy}
The local energy is mostly computed via the integration of responses values of pairs of quadrature filters, which makes it sensitive to noise and may lead to glitch artifact.
This is due to that the quadrature filters use fixed bases.
Although fixed bases are suitable for most types of signals, they are not adaptable to the input signals.
Hence, in this paper, we would like to search for an appropriate transform domain with data-driven bases to represent the signal and compute the corresponding local energy.

The transform domain is embedded in a Hilbert space $\mathbf{H}$ and composed of a set of complete orthogonal bases $\{\mathbf{v}_n\}_{n=1}^N, \mathbf{v}_n\in \mathbb{R}^N $, where $N$ is the dimension of $\mathbf{H}$.
Note that the wavelet transform can be used in the multiscale analysis naturally because the wavelet base is scalable, however, our method is data-driven, scaling of the bases will lose the complete and orthogonal property of the bases.
Therefore, for a 2D image $f(x,y)$, to acquire its information on different frequencies, firstly, we choose image patches around each pixel with different sizes $S_1, S_2, ..., S_K$ that are sorted in ascending order.
We denote these patches by $P_{1}, P_{2},...,P_{K}$, where $P_{k}\in\mathbb R^{\sqrt{S_k}\times \sqrt{S_k}}, S_k \in [S_1, S_K]$.
Then, at each target pixel, all of the patches are re-sampled to the size of $S_1$:
\begin{equation}\label{eq:downsample}
    P'_{k} = P_{k}(x,y)\downarrow_{\frac{S_k}{S_1}},
\end{equation}
where $\downarrow$ means the downsampling operation, $\frac{S_k}{S_1}$ is the downsampling ratio and $P'_{k}$ is the $k_{th}$ new patch after downsampling.
In this way, $P'_{1}$ (the same as $P_{1}$) contains the high-frequency information because it describes the local information within a small range.
Additionally, as $k$ increases, $P'_{k}$ contains lower and lower frequency information because the downsampling process has removed high-frequency information gradually.

The re-sampling process is similar to the calculation of phase congruency by using wavelet. The wavelet-based methods pick up the low-, mid- and high-frequency information according to the Log-Gabor filter on the frequency domain, while we extract different frequency components directly from the signal in the spatial domain.

Suppose the ideal transform domain is composed of a set of complete orthogonal bases $\{\mathbf{v}_n\}_{n=1}^{S_1}, \mathbf{v}_n\in \mathbb{R}^{S_1}$, then the edge detection process based on the local energy is computed as follows.

First, remove the DC component of each patch surrounded at each target pixel by subtracting their mean value respectively:
\begin{equation}\label{Xp}
   X_{k}  = P'_{k} - \bar P'_{k}.
\end{equation}

Next, vectorize these patches $\{X_{k}\}$ as $\{\mathbf{x}_{k}\}_{{k}=1}^{K}, \mathbf {x}_{k}\in \mathbb R^{S_1}$, and project all the vectors that carry different frequency information on $\{\mathbf{v}_n\}$:
\begin{equation}\label{eq:projection}
    \mathbf y_k = [\mathbf{x}_{k}^T\mathbf v_1,\mathbf{x}_{k}^T\mathbf v_2, ..., \mathbf{x}_{k}^T\mathbf v_{S_1}]^T,
\end{equation}
where $\mathbf y_k=[y_k^1,y_k^2,...,y_k^{S_1}]^T$ means the projection term of the vector from the $k_{th}$ scale, and $y_k^s = \mathbf{x}_{k}^T\mathbf v_s$ is the $s_{th}$ element of $\mathbf y_k$.

Then the summation of local energy can be expressed as follows:
\begin{equation}\label{eq:pcaenergy}
    E(x,y)=\sqrt{(\Sigma_k y_k^1)^2 + (\Sigma_k y_k^2)^2 + ... +(\Sigma_k y_k^{S_1})^2}.
\end{equation}

And the corresponding local amplitude can be computed as:
\begin{equation}\label{eq:pcaamplitude}
    \Sigma_k A_k(x,y) = \Sigma_k \sqrt{(y_k^1)^2+(y_k^2)^2+ ...+ (y_k^{S_1})^2}.
\end{equation}

\textbf{Definition}: The spectrum congruency based on the multiscale patches is defined as follows:
\begin{equation}\label{eq:sc}
    SC(x,y) = \frac{\lfloor E(x,y)-T\rfloor}{\Sigma_k A_k}.
\end{equation}

The term $T$ in \eq(\ref{eq:sc}) is a constant to compensate the influence of noise.
It can be determined either by the mean value of energy response, or by a fixed threshold given by the users.
In this paper, we use $T=\alpha\mu$ to estimate the noise energy, where $\mu$ stands for the mean value of energy response, and $\alpha$ is a given constant.

\subsection{Spectrum Congruency for Edge detection}
The spectrum congruency can be regarded as the edge strength map that reflects the probability of a pixel being the edge.
However, the pixels near the edges are also assigned a lot of energy, which means they have a very high probability of being edge pixels, thus degrading the accuracy in edge location.
Therefore, we take advantage of the non-maximum suppression algorithm \cite{Canny86} to thin the edges.
The edge maps after thinning is denoted $ETM$.
Algorithm.\ref{algorithm} describes how to obtain the $SC$ and $ETM$ in detail.

\begin{algorithm}
\caption{Spectrum Congruency of Multiscale Local Patches for Edge Detection}
\SetAlgoLined
\vspace{0.1cm}
\textbf{Input}: image $f(x,y)$, PatchSize=\{$S_1, S_2, S_3, ..., S_K$\};\\
{\textbf{Output}: spectrum congruency $SC$ and edge thinning map $ETM$;\\}
\algrule
\textbf{foreach} pixel $(x,y)$ in $f(x,y)$ \textbf{do} \\
\qquad 1. Extract $K$ patches centered at $(x,y)$: $P_{1}(x,y)$,\\
\qquad\quad $P_{2}(x,y)$, $P_{3}(x,y)$,..., $P_{K}(x,y)$\\
\qquad 2. Remove the DC components of each patch\\
\qquad  \textbf{foreach} $S_k$ in PatchSize \textbf{do}\\
\qquad \qquad 3. Downsample $P_{k}(x, y)$ to form a new patch \\
\qquad \qquad \quad $P_{S_knew}(x, y)$ with the same size of $S_1$\\
\qquad \qquad 4. Project each patch $P_{S_knew}(x, y)$ to a set of\\
\qquad \qquad \quad complete and orthogonal bases $\{\mathbf v\}_n$;\\
\qquad \qquad 5. Compute $E_k$ (\eq(\ref{eq:pcaenergy})) and $A_k$ respectively \\
\qquad \qquad \quad according to \eq(\ref{eq:pcaamplitude})\\
\qquad \textbf{endfor}\\
\qquad 6. Compute the spectrum congruency $SC(x,y)$ via \\
\qquad \quad (\eq(\ref{eq:sc}))\\
\textbf{endfor}\\
7. Apply NMS algorithm to $SC$ obtain the edge thinning \\
\quad map $ETM$\\
{\Return} $SC$, $ETM$\\
\label{algorithm}
\end{algorithm}

\begin{figure*}
  \centering
  {\includegraphics[width=1\textwidth]{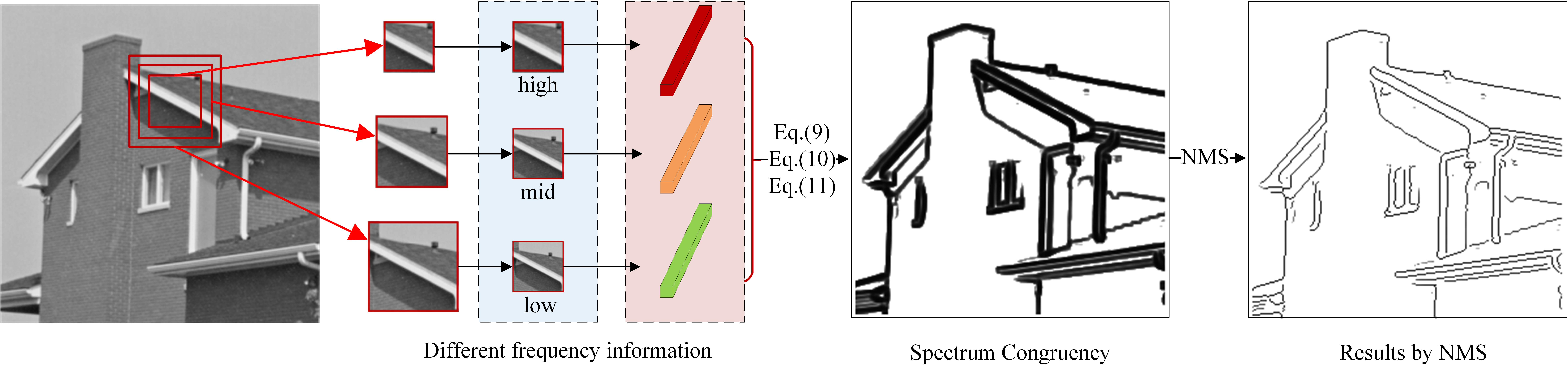}}\\
  \caption{Flowchart of the proposed method, three patches of different sizes are selected at a target point }
  \label{fig:flowchart}
\end{figure*}

Figure.\ref{fig:flowchart} depicts the whole process of the proposed spectrum congruency method.
We can see that the edges in the edge strength map obtained by \eq(\ref{eq:sc}) are very clear, and nearly all edges are detected.
However, the edge features are too wide to provide precise localization.
The non-maximum suppression algorithm is applied as the final step to provide a refined edge thinning map.

Compared with the phase congruency, the proposed spectrum congruency is based on the local energy of multiscale patches and does not need to consider the influence of the filter orientations and the frequency spread as the phase congruency does.
This avoids the integration of values from multiple orientations, thus eliminating the glitch artifact and providing a much simpler way to measure the edge strength.
Besides, there is no need of the weighting function to reduce the spurious response and the proposed method is highly robust to noise.
This is because the combination of low-, mid- and high- frequency information around each pixel helps to retain the most important information and remove the noise as well.

\subsection{Bases Selection}
\textbf{Theorem}:
Let $\{\mathbf{v}_n\}_{n=1}^N$, $\{\mathbf{u}_n\}_{n=1}^N$ be two arbitrary sets of complete orthogonal bases of a domain $\Omega \subset \mathbb{R}^N$.
For a set of vectors $\{\mathbf{x}_p\}_{p=1}^P, \mathbf{x}_p\in\Omega$, the energy and amplitude of its projection on $\{\mathbf{v}_n\}$ and $\{\mathbf{u}_n\}$ are the same:
\begin{equation*}
\left
\{\begin{matrix}
E_{\{\mathbf{v}_n\}} = E_{\{\mathbf{u}_n\}}\\
A_p(x)_{\{\mathbf{v}_n\}} = A_p(x)_{\{\mathbf{u}_n\}}
\end{matrix}.
\right.
\end{equation*}

\textbf{Proof}:
The projection of $\{\mathbf{x}_p\}$ on $\{\mathbf{v}_n\}$ and $\{\mathbf{u}_n\}$ are $\{\mathbf{y}_p\}$ and $\{\mathbf{z}_p\}$ respectively.
At each scale $p$, the projected vectors are:
\begin{equation*}\label{projection}
\left
\{\begin{matrix}
\mathbf y_p = [\mathbf x_p^T\mathbf v_1,\mathbf x_p^T\mathbf v_2, ..., \mathbf x_p^T\mathbf v_{N}]^T
\vspace{0.15cm}\\
\mathbf z_p = [\mathbf x_p^T\mathbf u_1,\mathbf x_p^T\mathbf u_2, ..., \mathbf x_p^T\mathbf u_{N}]^T,
\end{matrix}
\right.
\end{equation*}

The energy of the projected vector is computed according to Eq.(\ref{eq:pcaenergy}), the square of the energy is as follows:

\begin{equation}\label{eq:energyvn}
\begin{aligned}
E^2_{\{\mathbf{v}_n\}} = &(\mathbf x_1^T\mathbf v_1+\mathbf x_2^T\mathbf v_1+...+\mathbf x_P^T \mathbf v_1)^2 +\\
&  (\mathbf x_1^T\mathbf v_2+\mathbf x_2^T\mathbf v_2+...+\mathbf x_P^T \mathbf v_2)^2 + ... \\
&+ (\mathbf x_1^T\mathbf v_N+\mathbf x_2^T\mathbf v_N+...+\mathbf x_P^T \mathbf v_N)^2
\end{aligned}
\end{equation}

Let $(\mathbf x_1^T+\mathbf x_2^T+...+\mathbf x_P^T) =\mathbf x$, then \eq(\ref{eq:energyvn}) can be rewritten as:

\begin{equation}\label{eq:energyre}
    E^2_{\{\mathbf{v}_n\}} = (\mathbf x^T \mathbf v_1)^2+(\mathbf x^T \mathbf v_2)^2+...+(\mathbf x^T \mathbf v_N)^2
\end{equation}

\begin{figure}[ht]
  \includegraphics[width=0.5\textwidth]{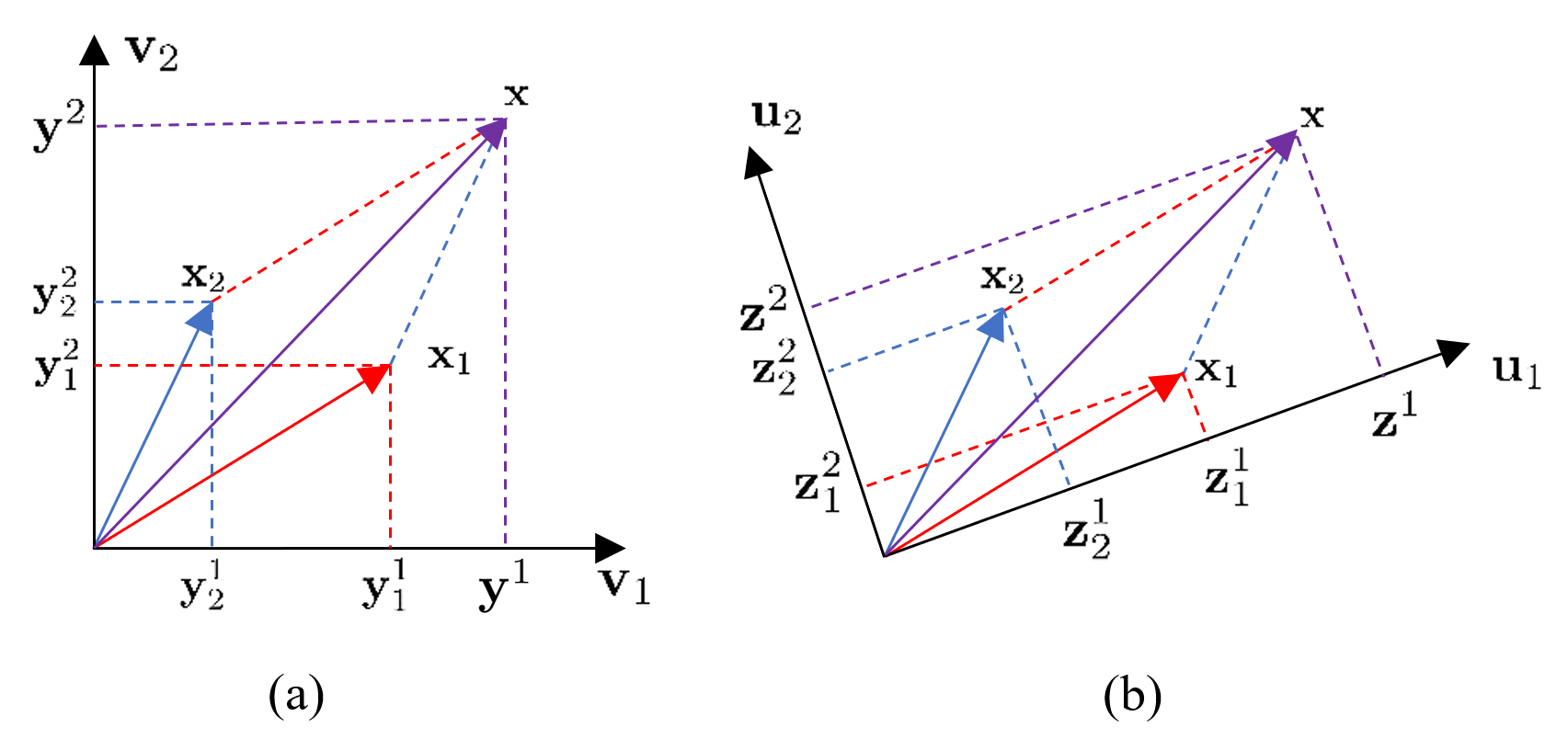}\\
  \caption{The sketch map of the projection of $\{\mathbf{x}_p\}_{p=1}^3$ on two complete orthogonal bases $\{\mathbf{v}_n\}_{n=1}^2$ and $\{\mathbf{u}_n\}_{n=1}^2$}
  \label{fig:projection}
\end{figure}

As illustrated in the sketch map \f\ref{fig:projection}, the projection of $\mathbf x$ on each base of $\{\mathbf{v}_n\}$ is: $\mathbf x^T \mathbf v_n = \mathbf y^n$, then the \emph{r.h.s} of \eq(\ref{eq:energyre}) is in fact the magnitude of the corresponding vector $\mathbf x$:

\begin{equation*}
(\mathbf y^1)^2+(\mathbf y^2)^2+...+(\mathbf y^N)^2 = \|\mathbf x\|^2
\end{equation*}

Similarly, when using the bases $\{\mathbf{u}_n\}$, the energy is $E^2_{\{\mathbf{u}_n\}} = (\mathbf z^1)^2 +(\mathbf z^2)^2+...+(\mathbf z^N)^2 = \|\mathbf x\|^2$.
Therefore, the energy of the projected vectors is actually the same regardless of the bases:
\begin{equation*}
E_{\{\mathbf{v}_n\}} = E_{\{\mathbf{u}_n\}}.
\end{equation*}

The amplitude of the projected vector $\mathbf y_p$ is computed according to Eq.(\ref{eq:pcaenergy}):
\begin{equation*}
A_p(x)_{\{\mathbf{v}_n\}} = \sqrt{(\mathbf x_p\mathbf v_1)^2+(\mathbf x_p\mathbf v_2)^2+...+(\mathbf x_p\mathbf v_N)^2}.
\end{equation*}

Due to that $\{\mathbf{v}_n\}$ is complete and orthogonal, $A_p(x)_{\{\mathbf{v}_n\}}$ is actually the $\mathfrak{L}_2$ norm of $\mathbf x_p$, \emph{i.e.}, $A_p(x)_{\{\mathbf{v}_n\}} = \|\mathbf x_p\|$, hence we have:
\begin{equation*}
A_p(x)_{\{\mathbf{v}_n\}} = A_p(x)_{\{\mathbf{u}_n\}}.
\end{equation*}
$\hfill\blacksquare$


According to this theorem, the spectrum congruency is unaffected by the bases as long as the transform domain is embedded in a Hilbert space $\mathbf H$, where the bases are all complete and orthogonal.
In other words, the spectrum congruency is all the same for all sets of complete orthogonal bases $\{\mathbf v_n\}_{n=1}^N$.
Hence, in our case, to facilitate the computation, we use the column vectors of the identity matrix $I_d \in \mathbb R^{S_1\times S_1}$ as the bases.
Thus, the energy and amplitude of the whole image can be calculated in linear time.

\section{Experiment and Analysis}
\label{sec:experiment}
\subsection{Experiment Data and Settings}
To verify the effectiveness of our proposed method, we test our approach on both noise-free images and noisy images.
The data used in this section are from the public dataset, including the BSDS500 dataset \cite{amfm_pami2011}.

In terms of the estimated noise energy $T=\alpha\mu$ in \eq(\ref{eq:sc}).
In our experiments, $\alpha$ is tunable.
Empirically, we set $\alpha=0.5$ for noise-free images and $1\leq\alpha\leq2.5$ for noisy images according to the noise deviation.
Three scales of patches are used in this paper, \emph{i.e. }$K=3$.
The choice of patch size is discussed in Sect.\ref{sec:patchsize}.

In Sect.\ref{sec:noisy}, we test the noise-immunity of the proposed method on noisy cases.
Noisy images are generated by adding Gaussian white additive noise to the original noise-free images.
We compare the spectrum congruency with the Log-Gabor-based and monogenic signal-based phase congruency.
In addition, we compare the edge-thinning maps of spectrum congruency with three state-of-the-art methods: the Canny edge detector (CED) \cite{Canny86}, the scale multiplication Canny edge detector (SMED) \cite{Bao2005Canny} and the detector based on isotropic and anisotropic Gaussian kernel (IAGK) \cite{2012Noise}.
The \emph{figure of merit} is considered as an objective measurement to evaluate the performance of the different methods under noisy cases:
\begin{equation}\label{FOM}
    FOM = \dfrac{1}{\max\{N_{gt}, N_{det}\}}\sum_{k=1}^{N_{det}}\dfrac{1}{1+\beta d^2(k)},
\end{equation}
where $N_{gt}$ represents the number of pixels of the groundtruth, ${N_{det}}$ stands for number of pixels of the edges detected by the edge detectors, $d(k)$ means the distance between the $k$-th real edge and the detected edge, $\alpha$ is a constant, normally, $\beta = 1/9$.

At last, we compare our method with the aforementioned state-of-the-art edge detectors on the standard benchmark dataset in Sect.\ref{sec:comparison}.
To evaluate the performance of these methods on standard benchmark dataset, we compute the precision, recall and F-measure:
\begin{equation}\label{eq:evaluation}
    \left\{
      \begin{array}{l}
        precision = \dfrac{N_{corr}}{N_{gt}}  \\
        recall = \dfrac{N_{corr}}{N_{det}} \vspace{0.2cm} \qquad\qquad\qquad ,\\
        F = \dfrac{2*precision*recall}{precision+recall}
      \end{array}
    \right.
\end{equation}
where ${N_{corr}}$ means the edges that are correctly detected,
There are two common ways of determining the optimal threshold of F-measure.
The first one is applying a fixed threshold to all the edge strength maps, which is called the optimal dataset scale (ODS) threshold.
The second one is to use an threshold to the edge strength maps individually, which is named the optimal image scale (OIS).
Details are discussed and analyzed as follows.

\subsection{Analysis of Patch Size}
\label{sec:patchsize}
\begin{figure}[htp]
  \centering
  \hspace{-0.3cm}
  {
  \setlength{\fboxrule}{0.01cm}
  \setlength{\fboxsep}{0cm}
  \fbox{\includegraphics[width=0.15\textwidth]{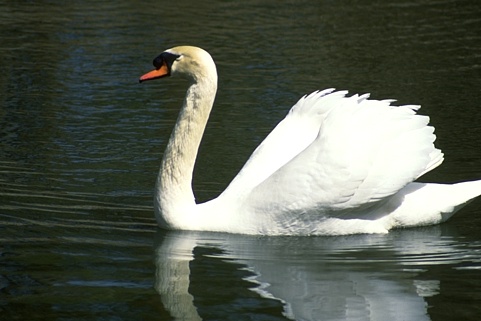}}}
  \hspace{-0.3cm}
  {
  \setlength{\fboxrule}{0.01cm}
  \setlength{\fboxsep}{0cm}
  \fbox{\includegraphics[width=0.15\textwidth]{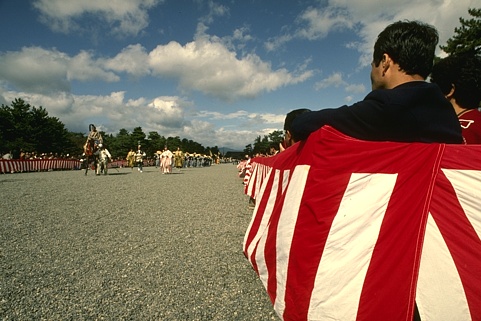}}}
  \hspace{-0.3cm}
  {
  \setlength{\fboxrule}{0.01cm}
  \setlength{\fboxsep}{0cm}
  \fbox{\includegraphics[width=0.15\textwidth]{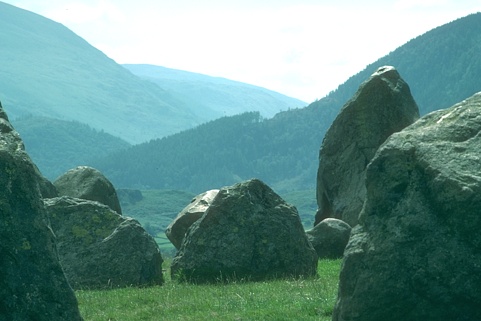}}}
  \caption{Images for analyzing the patch size}
  \label{patchsizeori}
\end{figure}

The patch size is an essential factor in our method.
Theoretically, the patch size can range from a single pixel $1\times1$ to the whole image $M\times N$. However, the patch size can not be too large in reality, or the localization of edge features will be inaccurate.
To test and analyze the impact of patch sizes, we study the patch size in two aspects: 1) the selection of the scales, or the gap size between two scale from fine to coarse patches; 2) with a fixed gap size, the size of patches from small to large.
Figure \ref{patchsizeori} displays the original three test images in this section.

\subsubsection{The selection of the scales}
The patch sizes reflect the bandwidths in the transform domain.
we choose different patch scales from fine to coarse with increasing gap size:
\begin{description}
  \item[Scale Case 1:] \quad\qquad $S_1=3\times3$, $S_2=5\times5$, $S_3=7\times7$;
  \item[Scale Case 2:] \quad\qquad $S_1=3\times3$, $S_2=7\times7$, $S_3=11\times11$;
  \item[Scale Case 3:] \quad\qquad $S_1=3\times3$, $S_2=11\times11$, $S_3=19\times19$.
\end{description}

\begin{figure*}[ht]
\centering
  \hspace{-0.3cm}
  \subfigure[]{
  \setlength{\fboxrule}{0.001cm}
  \setlength{\fboxsep}{0cm}
  \fbox{\includegraphics[width=0.15\textwidth]{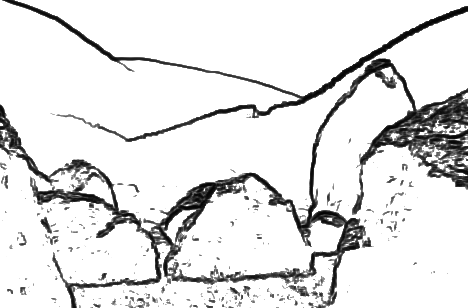}}}
  \hspace{-0.3cm}
  \subfigure[]{
  \setlength{\fboxrule}{0.001cm}
  \setlength{\fboxsep}{0cm}
  \fbox{\includegraphics[width=0.15\textwidth]{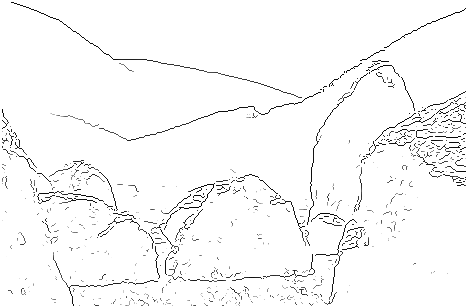}}}
  \hspace{-0.3cm}
  \subfigure[]{
  \setlength{\fboxrule}{0.001cm}
  \setlength{\fboxsep}{0cm}
  \fbox{\includegraphics[width=0.15\textwidth]{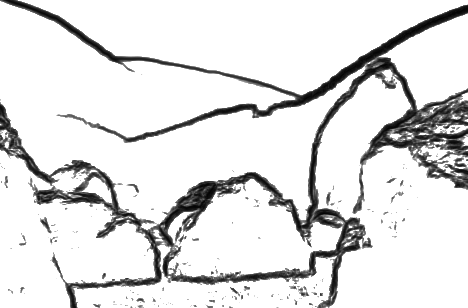}}}
  \centering
  \hspace{-0.3cm}
  \subfigure[]{
  \setlength{\fboxrule}{0.001cm}
  \setlength{\fboxsep}{0cm}
  \fbox{\includegraphics[width=0.15\textwidth]{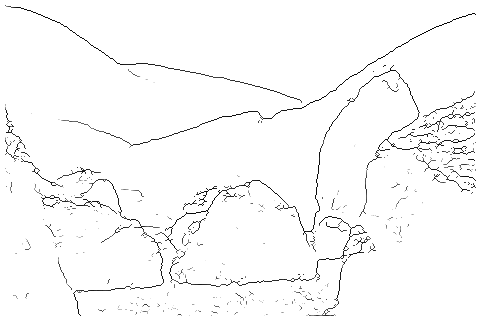}}}
  \hspace{-0.3cm}
  \subfigure[]{
  \setlength{\fboxrule}{0.001cm}
  \setlength{\fboxsep}{0cm}
  \fbox{\includegraphics[width=0.15\textwidth]{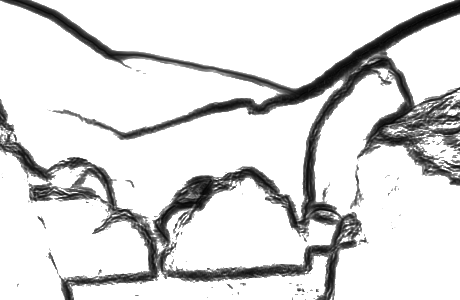}}}
  \hspace{-0.3cm}
  \subfigure[]{
  \setlength{\fboxrule}{0.001cm}
  \setlength{\fboxsep}{0cm}
  \fbox{\includegraphics[width=0.15\textwidth]{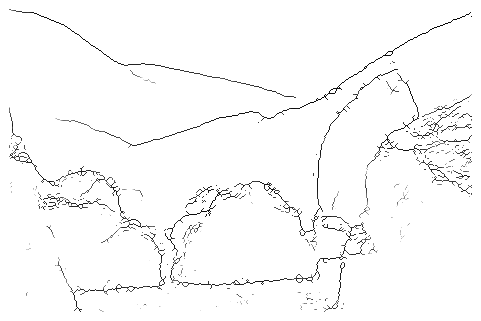}}}
  \caption{From left to right, every two columns are the edge strength maps and its corresponding thinning result; the results are obtained by using patch sizes from fine to coarse (Scale Case1 to Scale Case3) respectively.}
  \label{largepatchsize}
\end{figure*}

Figure \ref{largepatchsize} demonstrates the spectrum congruency and the thinning results of the \emph{mountain} image.
The results from left to right are obtained using patches with size from Scale Case 1 to Scale Case 3.
Clearly, with the increase of the patch size and the gap size, the extracted edge feature is broader and broader from (a) to (e), which results in severe glitch artifact after thinning.
See (f) as an example.
The reason is that, with an oversize patch, pixels far away from the edges will also be included in the big patch.
This means that the target/center pixel of the big patch is also assigned some energy from the edge pixel.
Hence, the patch size should be controlled within an appropriate range.
In our following experiments, the largest patch size is $11\times11$, and the gap size is no more than $4$.

\subsubsection{The selection of patch size}

Next, we would like to study that, within a suitable gap size, how the fine and coarse patch size would affect the spectrum congruency and the detected edge feature.
Therefore, three cases are considered here:

\begin{description}
  \item[Patchsize Case 1:] \qquad\qquad\quad$S_1=3\times3$, $S_2=5\times5$, $S_3=7\times7$;
  \item[Patchsize Case 2:] \qquad\qquad\quad $S_1=5\times5$, $S_2=7\times7$, $S_3=9\times9$;
  \item[Patchsize Case 3:] \qquad\qquad\quad $S_1=7\times7$, $S_2=9\times9$, $S_3=11\times11$.
\end{description}

\begin{figure*}[htp]
  \centering
  \hspace{-0.3cm}
  \setlength{\fboxrule}{0.001cm}
  \setlength{\fboxsep}{0cm}
  \fbox{\includegraphics[width=0.15\textwidth]{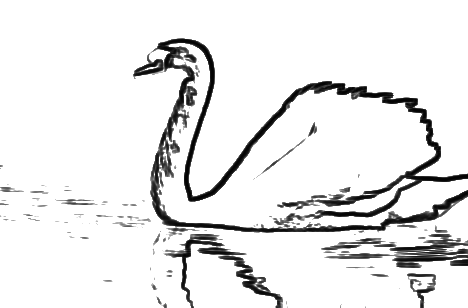}}
  \hspace{-0.3cm}
  \setlength{\fboxrule}{0.001cm}
  \setlength{\fboxsep}{0cm}
  \fbox{\includegraphics[width=0.15\textwidth]{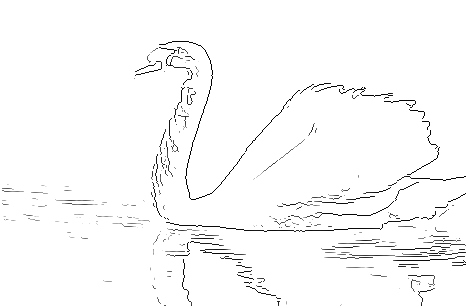}}
  \hspace{-0.3cm}
  \setlength{\fboxrule}{0.001cm}
  \setlength{\fboxsep}{0cm}
  \fbox{\includegraphics[width=0.15\textwidth]{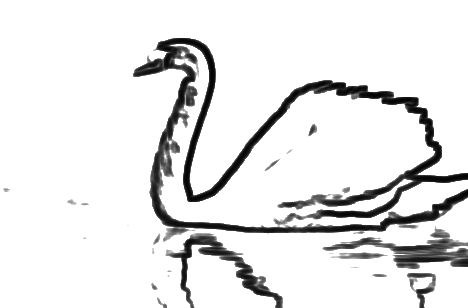}}
  \hspace{-0.3cm}
  \setlength{\fboxrule}{0.001cm}
  \setlength{\fboxsep}{0cm}
  \fbox{\includegraphics[width=0.15\textwidth]{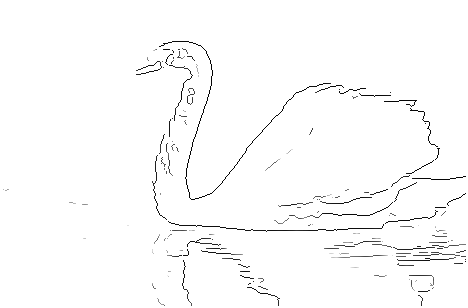}}
  \hspace{-0.3cm}
  \setlength{\fboxrule}{0.001cm}
  \setlength{\fboxsep}{0cm}
  \fbox{\includegraphics[width=0.15\textwidth]{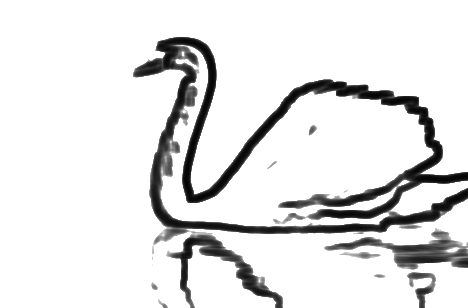}}
  \hspace{-0.3cm}
  \setlength{\fboxrule}{0.001cm}
  \setlength{\fboxsep}{0cm}
  \fbox{\includegraphics[width=0.15\textwidth]{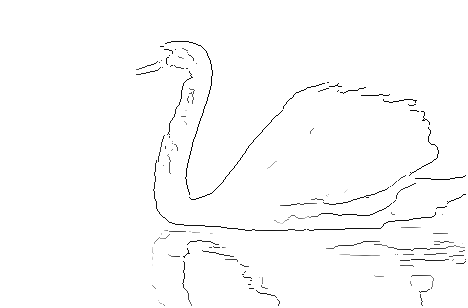}}\\[1pt]
  \hspace{-0.3cm}
  \setlength{\fboxrule}{0.001cm}
  \setlength{\fboxsep}{0cm}
  \fbox{\includegraphics[width=0.15\textwidth]{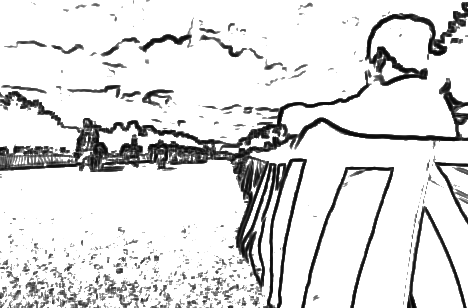}}
  \hspace{-0.3cm}
  \setlength{\fboxrule}{0.001cm}
  \setlength{\fboxsep}{0cm}
  \fbox{\includegraphics[width=0.15\textwidth]{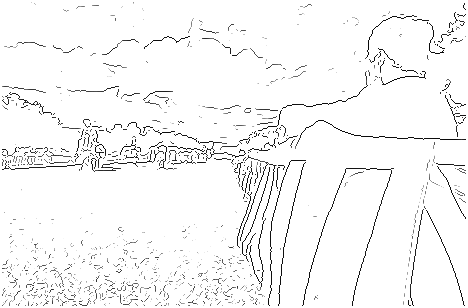}}
  \hspace{-0.3cm}
  \setlength{\fboxrule}{0.001cm}
  \setlength{\fboxsep}{0cm}
  \fbox{\includegraphics[width=0.15\textwidth]{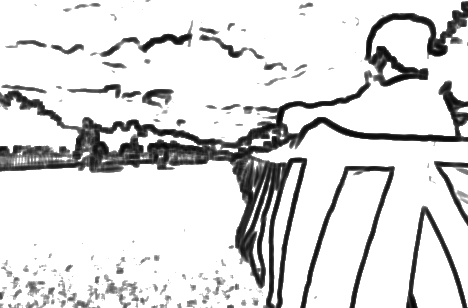}}
  \hspace{-0.3cm}
  \setlength{\fboxrule}{0.001cm}
  \setlength{\fboxsep}{0cm}
  \fbox{\includegraphics[width=0.15\textwidth]{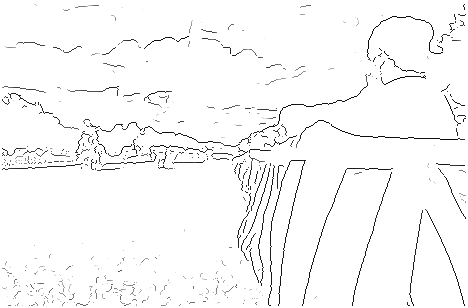}}
  \hspace{-0.3cm}
  \setlength{\fboxrule}{0.001cm}
  \setlength{\fboxsep}{0cm}
  \fbox{\includegraphics[width=0.15\textwidth]{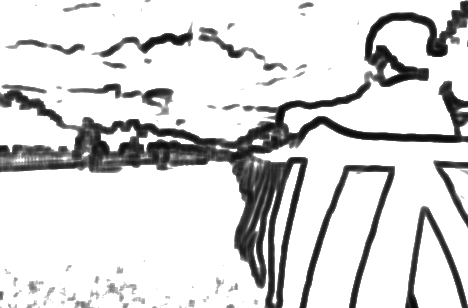}}
  \hspace{-0.3cm}
  \setlength{\fboxrule}{0.001cm}
  \setlength{\fboxsep}{0cm}
  \fbox{\includegraphics[width=0.15\textwidth]{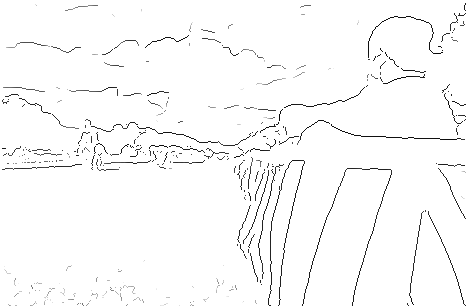}}\\[1pt]
  \hspace{-0.3cm}
  \setlength{\fboxrule}{0.001cm}
  \setlength{\fboxsep}{0cm}
  \fbox{\includegraphics[width=0.15\textwidth]{images/Mountains_PC357.png}}
  \hspace{-0.3cm}
  \setlength{\fboxrule}{0.001cm}
  \setlength{\fboxsep}{0cm}
  \fbox{\includegraphics[width=0.15\textwidth]{images/Mountains_PCthin357.png}}
  \hspace{-0.3cm}
  \setlength{\fboxrule}{0.001cm}
  \setlength{\fboxsep}{0cm}
  \fbox{\includegraphics[width=0.15\textwidth]{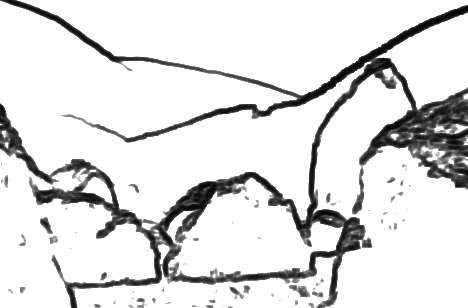}}
  \hspace{-0.3cm}
  \setlength{\fboxrule}{0.001cm}
  \setlength{\fboxsep}{0cm}
  \fbox{\includegraphics[width=0.15\textwidth]{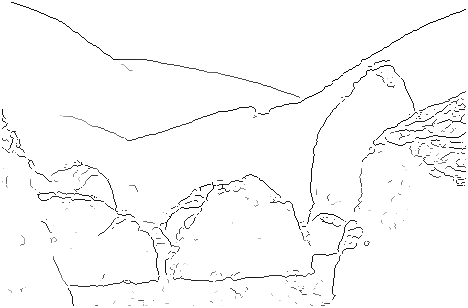}}
  \hspace{-0.3cm}
  \setlength{\fboxrule}{0.001cm}
  \setlength{\fboxsep}{0cm}
  \fbox{\includegraphics[width=0.15\textwidth]{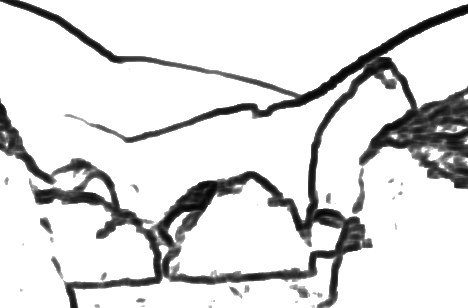}}
  \hspace{-0.3cm}
  \setlength{\fboxrule}{0.001cm}
  \setlength{\fboxsep}{0cm}
  \fbox{\includegraphics[width=0.15\textwidth]{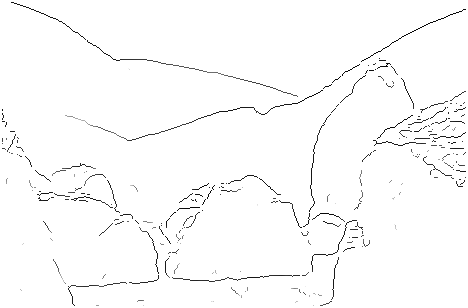}}\\
  \caption{from left to right, every two columns are the edge strength maps and its corresponding thinning result; the results are obtained by using patch sizes from fine to coarse (Patchsize Case1 to Patchsize Case3) respectively.}
  \label{patchsize}
\end{figure*}

The first to the third column of \f\ref{patchsize} shows the results obtained by the patch size of Patchsize Case 1 to Patchsize Case 3 respectively.
It can be seen clearly that as the patch size increases, the details are neglected gradually, and the contour of salient objects are more obvious and complete.
For example, in the \emph{goose} image, the fur on the neck of the goose and the water wave are detected by using the smallest patch size, while in Column.3, only a few of fur on the neck is preserved and the water wave is barely seen.
Additionally, in the \emph{parade} image, the stones on the street are less and less detected with the increase of patch size.
Moreover, in the \emph{mountain} image, the grass on the ground and cracks on the stone are mostly obtained by small patch size, while with big patch size, much fewer details are detected.

From the experimental results, we can conclude that small patch sizes are fit for extracting the edge feature of details in images, while big patch sizes are more suitable for detecting the contour of the salient objects in images.

\subsection{Experiments on Noisy Cases}
\label{sec:noisy}

\subsubsection{Spectrum Congruency {vs} Phase Congruency}
\begin{figure*}[ht]
  \centering
  \hspace{-0.3cm}\setlength{\fboxrule}{0.001cm}
  \setlength{\fboxsep}{0cm}
  \fbox  {\includegraphics[width=0.12\textwidth]{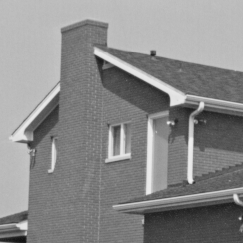}}
  \hspace{-0.3cm}\setlength{\fboxrule}{0.001cm}
  \setlength{\fboxsep}{0cm}
  \fbox  {\includegraphics[width=0.12\textwidth]{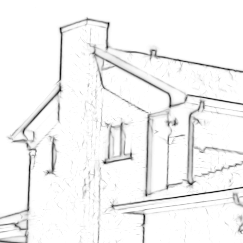}}
  \hspace{-0.3cm}\setlength{\fboxrule}{0.001cm}
  \setlength{\fboxsep}{0cm}
  \fbox  {\includegraphics[width=0.12\textwidth]{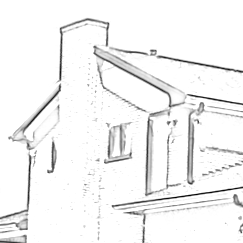}}
  \hspace{-0.3cm}\setlength{\fboxrule}{0.001cm}
  \setlength{\fboxsep}{0cm}
  \fbox  {\includegraphics[width=0.12\textwidth]{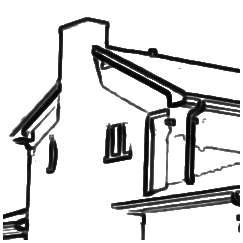}}
  \hspace{-0.3cm}\setlength{\fboxrule}{0.001cm}
  \setlength{\fboxsep}{0cm}
  \fbox  {\includegraphics[width=0.12\textwidth]{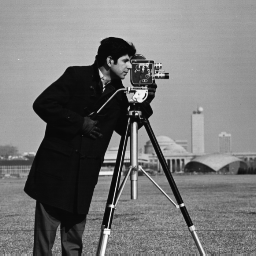}}
  \hspace{-0.3cm}\setlength{\fboxrule}{0.001cm}
  \setlength{\fboxsep}{0cm}
  \fbox  {\includegraphics[width=0.12\textwidth]{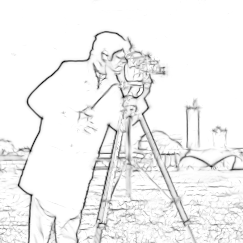}}
  \hspace{-0.3cm}\setlength{\fboxrule}{0.001cm}
  \setlength{\fboxsep}{0cm}
  \fbox  {\includegraphics[width=0.12\textwidth]{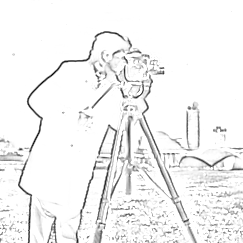}}  \hspace{-0.3cm}\setlength{\fboxrule}{0.001cm}
  \setlength{\fboxsep}{0cm}
  \fbox  {\includegraphics[width=0.12\textwidth]{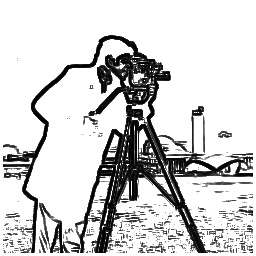}}\\[1pt]
  \hspace{-0.3cm}\setlength{\fboxrule}{0.001cm}
  \setlength{\fboxsep}{0cm}
  \fbox  {\includegraphics[width=0.12\textwidth]{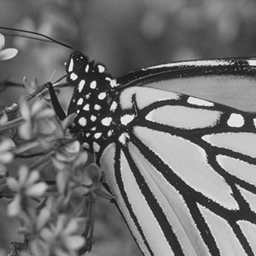}}
  \hspace{-0.3cm}\setlength{\fboxrule}{0.001cm}
  \setlength{\fboxsep}{0cm}
  \fbox  {\includegraphics[width=0.12\textwidth]{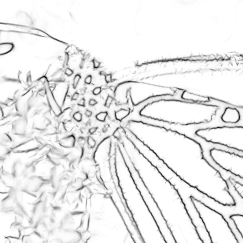}}
  \hspace{-0.3cm}\setlength{\fboxrule}{0.001cm}
  \setlength{\fboxsep}{0cm}
  \fbox  {\includegraphics[width=0.12\textwidth]{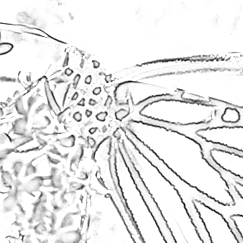}}
  \hspace{-0.3cm}\setlength{\fboxrule}{0.001cm}
  \setlength{\fboxsep}{0cm}
  \fbox  {\includegraphics[width=0.12\textwidth]{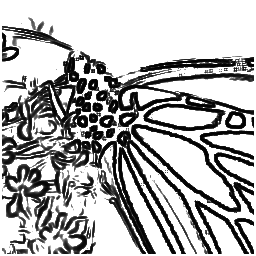}}
  \hspace{-0.3cm}\setlength{\fboxrule}{0.001cm}
  \setlength{\fboxsep}{0cm}
  \fbox  {\includegraphics[width=0.12\textwidth]{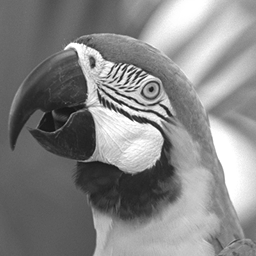}}
  \hspace{-0.3cm}\setlength{\fboxrule}{0.001cm}
  \setlength{\fboxsep}{0cm}
  \fbox  {\includegraphics[width=0.12\textwidth]{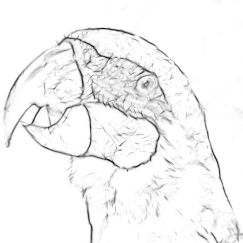}}
  \hspace{-0.3cm}\setlength{\fboxrule}{0.001cm}
  \setlength{\fboxsep}{0cm}
  \fbox  {\includegraphics[width=0.12\textwidth]{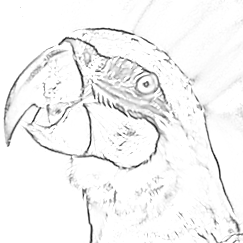}}
  \hspace{-0.3cm}\setlength{\fboxrule}{0.001cm}
  \setlength{\fboxsep}{0cm}
  \fbox  {\includegraphics[width=0.12\textwidth]{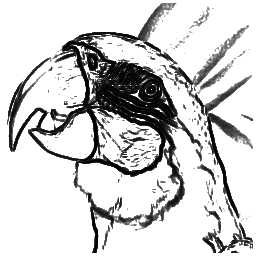}}\\[1pt]
  \hspace{-0.3cm}\setlength{\fboxrule}{0.001cm}
  \setlength{\fboxsep}{0cm}
  \fbox  {\includegraphics[width=0.12\textwidth]{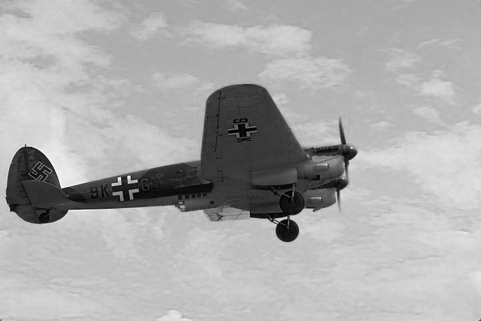}}
  \hspace{-0.3cm}\setlength{\fboxrule}{0.001cm}
  \setlength{\fboxsep}{0cm}
  \fbox  {\includegraphics[width=0.12\textwidth]{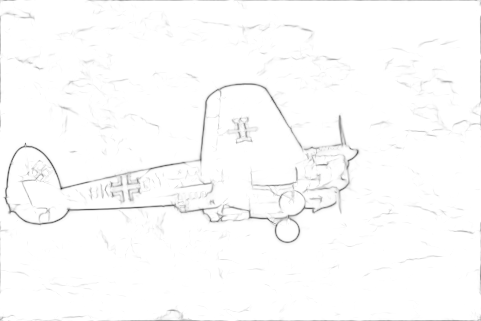}}
  \hspace{-0.3cm}\setlength{\fboxrule}{0.001cm}
  \setlength{\fboxsep}{0cm}
  \fbox  {\includegraphics[width=0.12\textwidth]{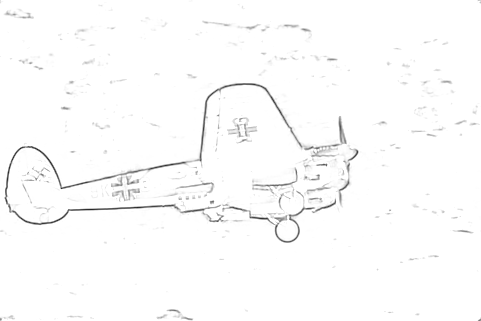}}
  \hspace{-0.3cm}\setlength{\fboxrule}{0.001cm}
  \setlength{\fboxsep}{0cm}
  \fbox  {\includegraphics[width=0.12\textwidth]{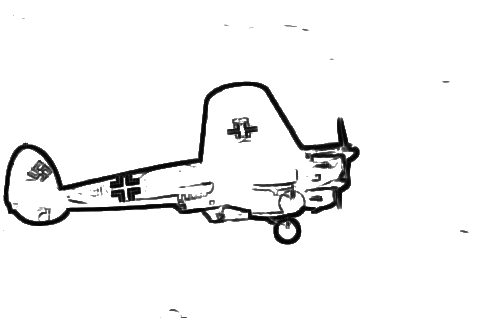}}
  \hspace{-0.3cm}\setlength{\fboxrule}{0.001cm}
  \setlength{\fboxsep}{0cm}
  \fbox  {\includegraphics[width=0.12\textwidth]{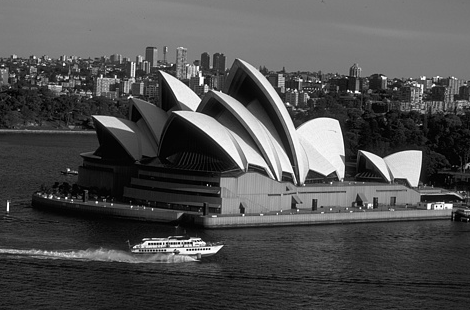}}
  \hspace{-0.3cm}\setlength{\fboxrule}{0.001cm}
  \setlength{\fboxsep}{0cm}
  \fbox  {\includegraphics[width=0.12\textwidth]{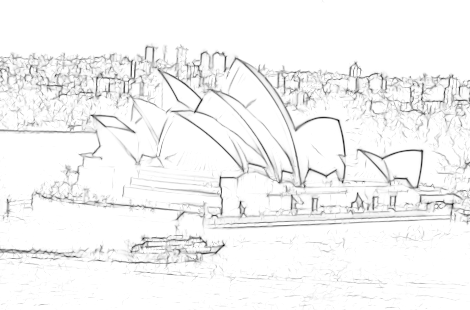}}
  \hspace{-0.3cm}\setlength{\fboxrule}{0.001cm}
  \setlength{\fboxsep}{0cm}
  \fbox  {\includegraphics[width=0.12\textwidth]{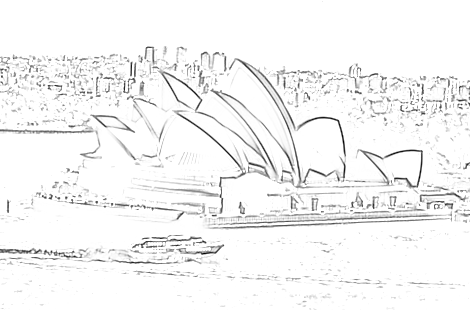}}
  \hspace{-0.3cm}\setlength{\fboxrule}{0.001cm}
  \setlength{\fboxsep}{0cm}
  \fbox  {\includegraphics[width=0.12\textwidth]{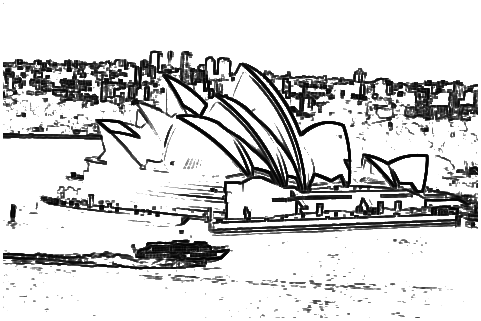}}\\[1pt]
  \caption{from left to right, every four columns are the original noisy free image and the edge strength maps based on Log-Gabor transform, monogenic signal and the proposed method respectively.}
  \label{fig:ori}
\end{figure*}

\begin{figure*}[htp]
  \centering
  \hspace{-0.3cm}\setlength{\fboxrule}{0.001cm}
  \setlength{\fboxsep}{0cm}
  \fbox{\includegraphics[width=0.12\textwidth]{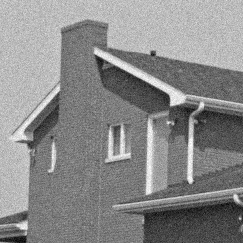}}
  \hspace{-0.3cm}\setlength{\fboxrule}{0.001cm}
  \setlength{\fboxsep}{0cm}
  \fbox{\includegraphics[width=0.12\textwidth]{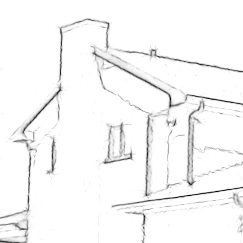}}
  \hspace{-0.3cm}\setlength{\fboxrule}{0.001cm}
  \setlength{\fboxsep}{0cm}
  \fbox{\includegraphics[width=0.12\textwidth]{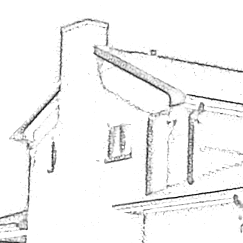}}
  \hspace{-0.3cm}\setlength{\fboxrule}{0.001cm}
  \setlength{\fboxsep}{0cm}
  \fbox{\includegraphics[width=0.12\textwidth]{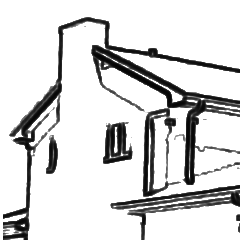}}
  \hspace{-0.3cm}\setlength{\fboxrule}{0.001cm}
  \setlength{\fboxsep}{0cm}
  \fbox{\includegraphics[width=0.12\textwidth]{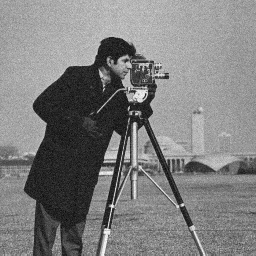}}
  \hspace{-0.3cm}\setlength{\fboxrule}{0.001cm}
  \setlength{\fboxsep}{0cm}
  \fbox{\includegraphics[width=0.12\textwidth]{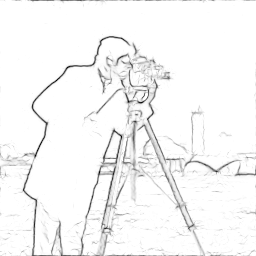}}
  \hspace{-0.3cm}\setlength{\fboxrule}{0.001cm}
  \setlength{\fboxsep}{0cm}
  \fbox{\includegraphics[width=0.12\textwidth]{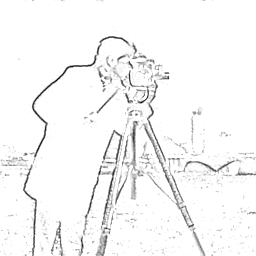}}
  \hspace{-0.3cm}\setlength{\fboxrule}{0.001cm}
  \setlength{\fboxsep}{0cm}
  \fbox{\includegraphics[width=0.12\textwidth]{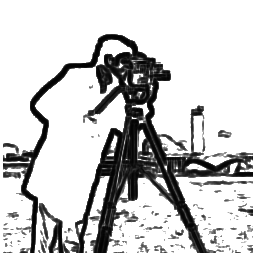}}\\[1pt]
  \hspace{-0.3cm}\setlength{\fboxrule}{0.001cm}
  \setlength{\fboxsep}{0cm}
  \fbox{\includegraphics[width=0.12\textwidth]{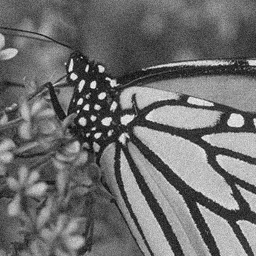}}
  \hspace{-0.3cm}\setlength{\fboxrule}{0.001cm}
  \setlength{\fboxsep}{0cm}
  \fbox {\includegraphics[width=0.12\textwidth]{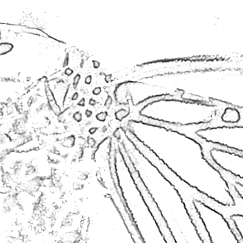}}
  \hspace{-0.3cm}\setlength{\fboxrule}{0.001cm}
  \setlength{\fboxsep}{0cm}
  \fbox{\includegraphics[width=0.12\textwidth]{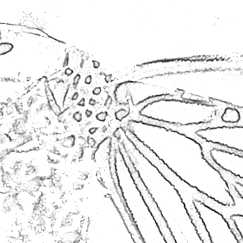}}
  \hspace{-0.3cm}\setlength{\fboxrule}{0.001cm}
  \setlength{\fboxsep}{0cm}
  \fbox{\includegraphics[width=0.12\textwidth]{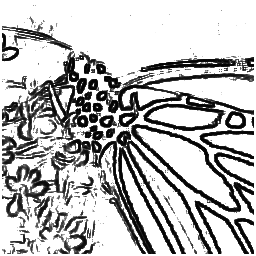}}
  \hspace{-0.3cm}\setlength{\fboxrule}{0.001cm}
  \setlength{\fboxsep}{0cm}
  \fbox{\includegraphics[width=0.12\textwidth]{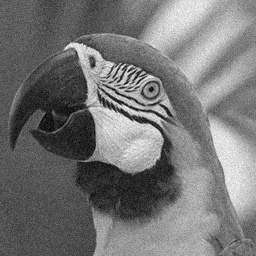}}
  \hspace{-0.3cm}\setlength{\fboxrule}{0.001cm}
  \setlength{\fboxsep}{0cm}
  \fbox{\includegraphics[width=0.12\textwidth]{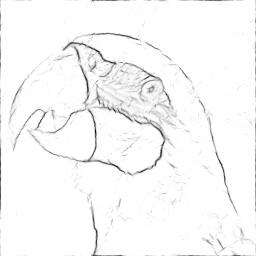}}
  \hspace{-0.3cm}\setlength{\fboxrule}{0.001cm}
  \setlength{\fboxsep}{0cm}
  \fbox{\includegraphics[width=0.12\textwidth]{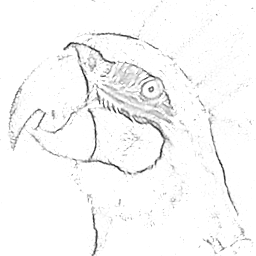}}
  \hspace{-0.3cm}\setlength{\fboxrule}{0.001cm}
  \setlength{\fboxsep}{0cm}
  \fbox{\includegraphics[width=0.12\textwidth]{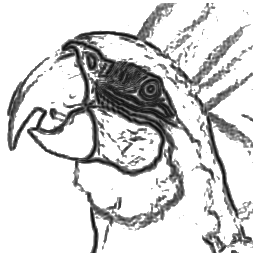}}\\[1pt]
  \hspace{-0.3cm}\setlength{\fboxrule}{0.001cm}
  \setlength{\fboxsep}{0cm}
  \fbox {\includegraphics[width=0.12\textwidth]{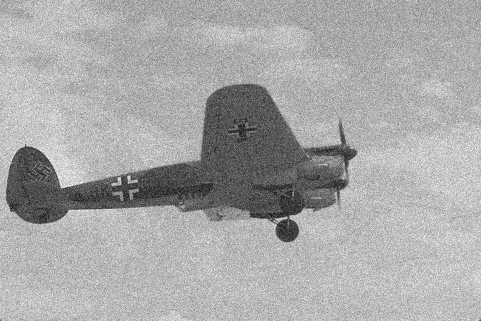}}
  \hspace{-0.3cm}\setlength{\fboxrule}{0.001cm}
  \setlength{\fboxsep}{0cm}
  \fbox {\includegraphics[width=0.12\textwidth]{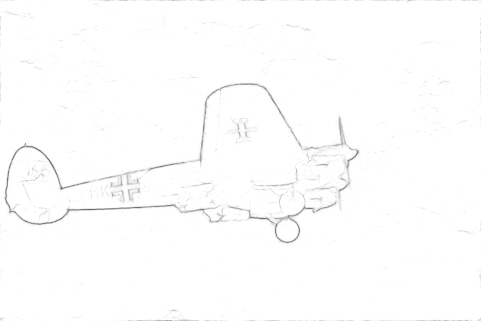}}
  \hspace{-0.3cm}\setlength{\fboxrule}{0.001cm}
  \setlength{\fboxsep}{0cm}
  \fbox{\includegraphics[width=0.12\textwidth]{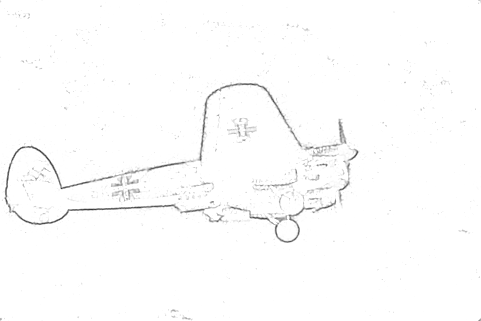}}
  \hspace{-0.3cm}\setlength{\fboxrule}{0.001cm}
  \setlength{\fboxsep}{0cm}
  \fbox {\includegraphics[width=0.12\textwidth]{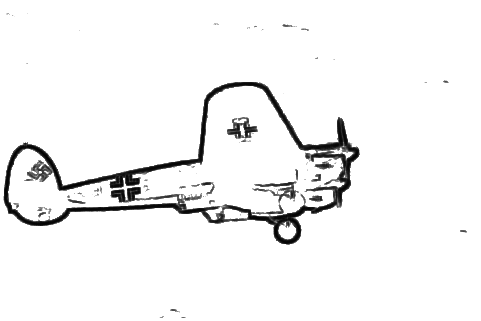}}
  \hspace{-0.3cm}\setlength{\fboxrule}{0.001cm}
  \setlength{\fboxsep}{0cm}
  \fbox {\includegraphics[width=0.12\textwidth]{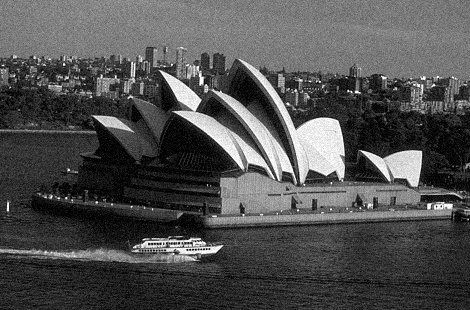}}
  \hspace{-0.3cm}\setlength{\fboxrule}{0.001cm}
  \setlength{\fboxsep}{0cm}
  \fbox  {\includegraphics[width=0.12\textwidth]{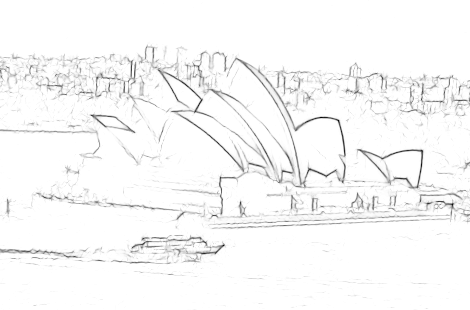}}
  \hspace{-0.3cm}\setlength{\fboxrule}{0.001cm}
  \setlength{\fboxsep}{0cm}
  \fbox  {\includegraphics[width=0.12\textwidth]{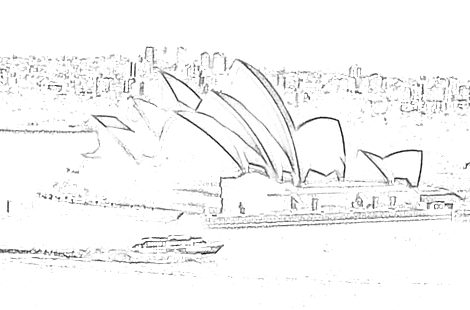}}
  \hspace{-0.3cm}\setlength{\fboxrule}{0.001cm}
  \setlength{\fboxsep}{0cm}
  \fbox  {\includegraphics[width=0.12\textwidth]{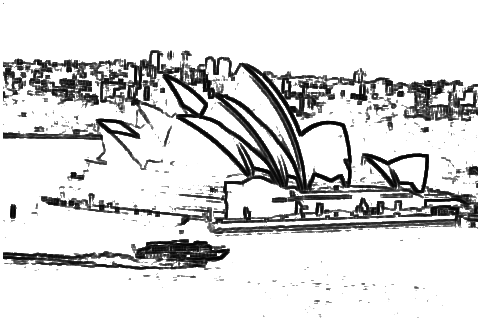}}\\[1pt]
  \vspace{-0.2cm}
  \caption{from left to right, every four columns are the noisy images, with the noise variance $\sigma = 10$, and the edge strength map based on Log-Gabor transform, monogenic signal and the proposed method respectively.}
  \label{fig:10}
 \end{figure*}

\begin{figure*}[htp]
  \centering
  \hspace{-0.3cm}\setlength{\fboxrule}{0.001cm}
  \setlength{\fboxsep}{0cm}
  \fbox{\includegraphics[width=0.12\textwidth]{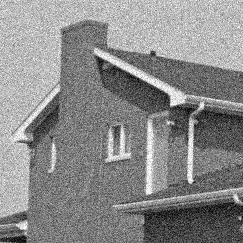}} \hspace{-0.3cm}\setlength{\fboxrule}{0.001cm}
  \setlength{\fboxsep}{0cm}
  \fbox{\includegraphics[width=0.12\textwidth]{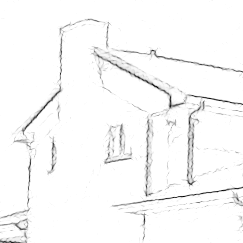}}
  \hspace{-0.3cm}\setlength{\fboxrule}{0.001cm}
  \setlength{\fboxsep}{0cm}
  \fbox{\includegraphics[width=0.12\textwidth]{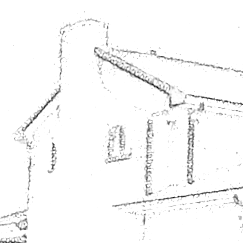}} \hspace{-0.3cm}\setlength{\fboxrule}{0.001cm}
  \setlength{\fboxsep}{0cm}
  \fbox{\includegraphics[width=0.12\textwidth]{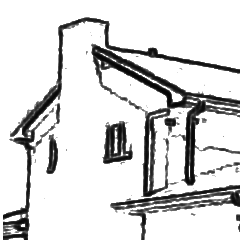}}
  \hspace{-0.3cm}\setlength{\fboxrule}{0.001cm}
  \setlength{\fboxsep}{0cm}
  \fbox{\includegraphics[width=0.12\textwidth]{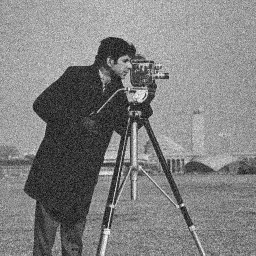}}
  \hspace{-0.3cm}\setlength{\fboxrule}{0.001cm}
  \setlength{\fboxsep}{0cm}
  \fbox{\includegraphics[width=0.12\textwidth]{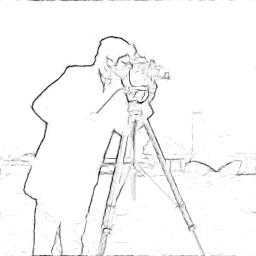}}
  \hspace{-0.3cm}\setlength{\fboxrule}{0.001cm}
  \setlength{\fboxsep}{0cm}
  \fbox{\includegraphics[width=0.12\textwidth]{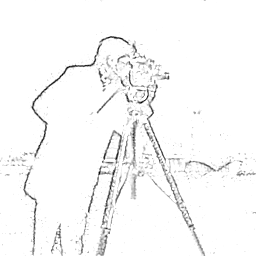}} \hspace{-0.3cm}\setlength{\fboxrule}{0.001cm}
  \setlength{\fboxsep}{0cm}
  \fbox{\includegraphics[width=0.12\textwidth]{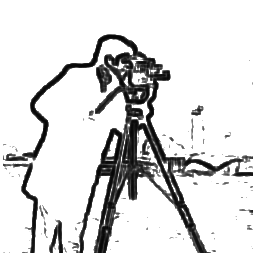}}\\[1pt]
  \hspace{-0.3cm}\setlength{\fboxrule}{0.001cm}
  \setlength{\fboxsep}{0cm}
  \fbox{\includegraphics[width=0.12\textwidth]{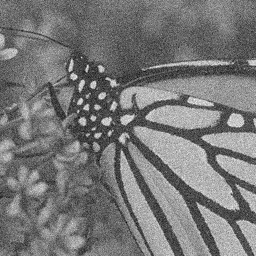}}
  \hspace{-0.3cm}\setlength{\fboxrule}{0.001cm}
  \setlength{\fboxsep}{0cm}
  \fbox{\includegraphics[width=0.12\textwidth]{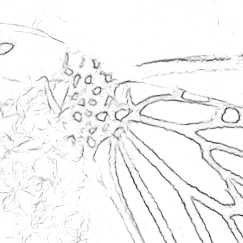}}
  \hspace{-0.3cm}\setlength{\fboxrule}{0.001cm}
  \setlength{\fboxsep}{0cm}
  \fbox{\includegraphics[width=0.12\textwidth]{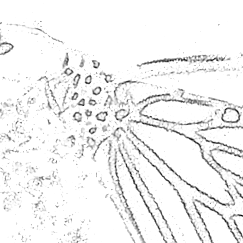}} \hspace{-0.3cm}\setlength{\fboxrule}{0.001cm}
  \setlength{\fboxsep}{0cm}
  \fbox{\includegraphics[width=0.12\textwidth]{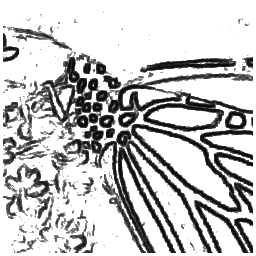}} \hspace{-0.3cm}\setlength{\fboxrule}{0.001cm}
  \setlength{\fboxsep}{0cm}
  \fbox{\includegraphics[width=0.12\textwidth]{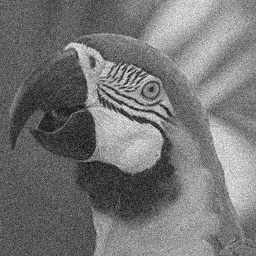}}
  \hspace{-0.3cm}\setlength{\fboxrule}{0.001cm}
  \setlength{\fboxsep}{0cm}
  \fbox{\includegraphics[width=0.12\textwidth]{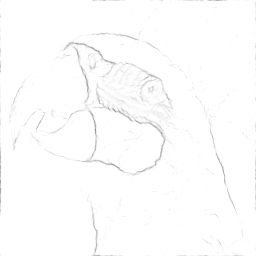}}
  \hspace{-0.3cm}\setlength{\fboxrule}{0.001cm}
  \setlength{\fboxsep}{0cm}
  \fbox{\includegraphics[width=0.12\textwidth]{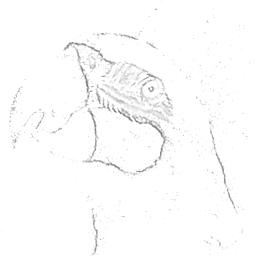}} \hspace{-0.3cm}\setlength{\fboxrule}{0.001cm}
  \setlength{\fboxsep}{0cm}
  \fbox{\includegraphics[width=0.12\textwidth]{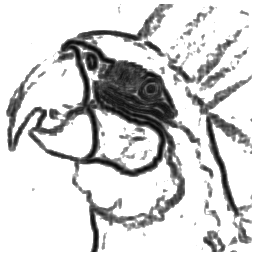}}\\[1pt]            \hspace{-0.3cm}\setlength{\fboxrule}{0.001cm}
  \setlength{\fboxsep}{0cm}
  \fbox{\includegraphics[width=0.12\textwidth]{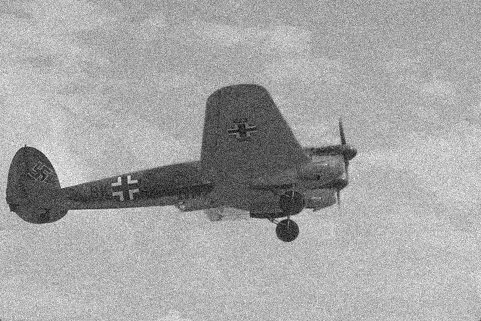}}
  \hspace{-0.3cm}\setlength{\fboxrule}{0.001cm}
  \setlength{\fboxsep}{0cm}
  \fbox{\includegraphics[width=0.12\textwidth]{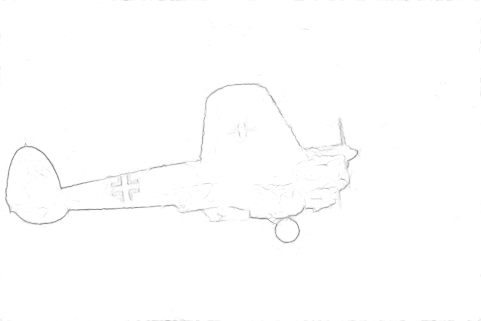}}
  \hspace{-0.3cm}\setlength{\fboxrule}{0.001cm}
  \setlength{\fboxsep}{0cm}
  \fbox{\includegraphics[width=0.12\textwidth]{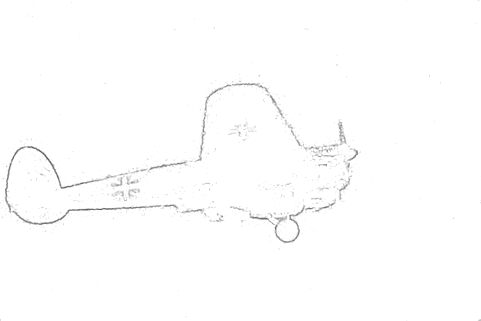}} \hspace{-0.3cm}\setlength{\fboxrule}{0.001cm}
  \setlength{\fboxsep}{0cm}
  \fbox{\includegraphics[width=0.12\textwidth]{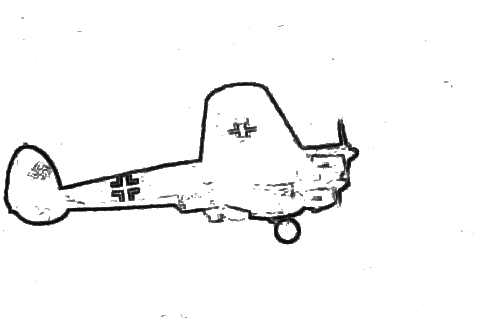}} \hspace{-0.3cm}\setlength{\fboxrule}{0.001cm}
  \setlength{\fboxsep}{0cm}
  \fbox{\includegraphics[width=0.12\textwidth]{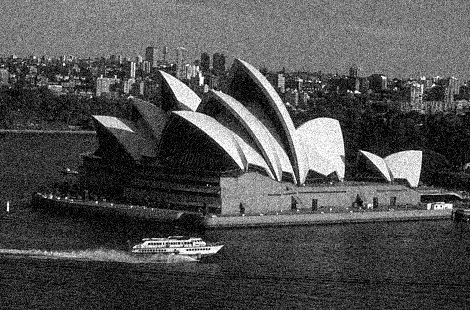}}
  \hspace{-0.3cm}\setlength{\fboxrule}{0.001cm}
  \setlength{\fboxsep}{0cm}
  \fbox{\includegraphics[width=0.12\textwidth]{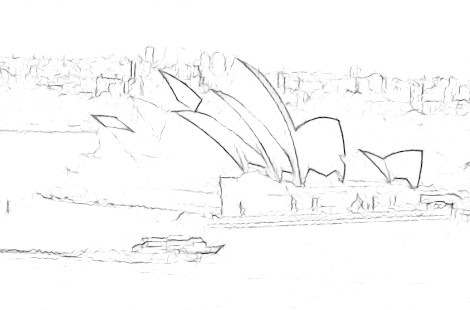}}
  \hspace{-0.3cm}\setlength{\fboxrule}{0.001cm}
  \setlength{\fboxsep}{0cm}
  \fbox{\includegraphics[width=0.12\textwidth]{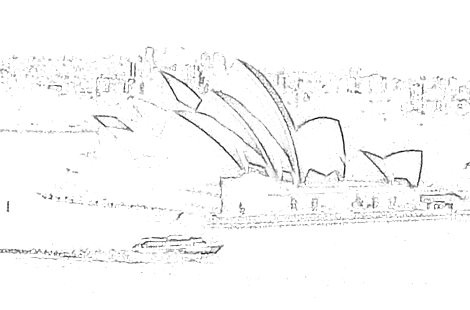}} \hspace{-0.3cm}\setlength{\fboxrule}{0.001cm}
  \setlength{\fboxsep}{0cm}
  \fbox{\includegraphics[width=0.12\textwidth]{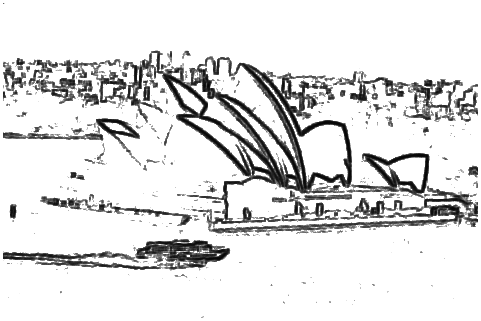}}\\[1pt]  \caption{from left to right, every four columns are the noisy images, with the noise variance $\sigma = 20$, and the edge strength map based on Log-Gabor transform, monogenic signal and the proposed method respectively.}
  \label{fig:20}
\end{figure*}

\begin{figure*}[htp]
  \centering
  \hspace{-0.3cm}\setlength{\fboxrule}{0.001cm}
  \setlength{\fboxsep}{0cm}
  \fbox{\includegraphics[width=0.12\textwidth]{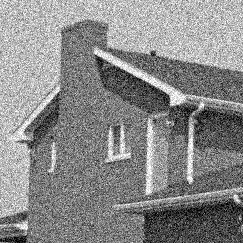}}
  \hspace{-0.3cm}\setlength{\fboxrule}{0.001cm}
  \setlength{\fboxsep}{0cm}
  \fbox{\includegraphics[width=0.12\textwidth]{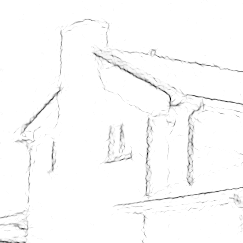}}
  \hspace{-0.3cm}\setlength{\fboxrule}{0.001cm}
  \setlength{\fboxsep}{0cm}
  \fbox{\includegraphics[width=0.12\textwidth]{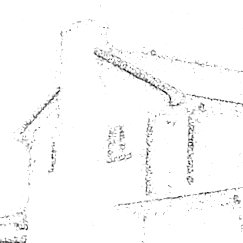}}
  \hspace{-0.3cm}\setlength{\fboxrule}{0.001cm}
  \setlength{\fboxsep}{0cm}
  \fbox{\includegraphics[width=0.12\textwidth]{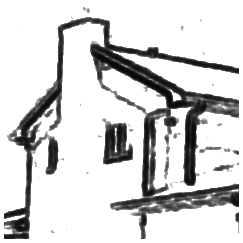}}
  \hspace{-0.3cm}\setlength{\fboxrule}{0.001cm}
  \setlength{\fboxsep}{0cm}
  \fbox{\includegraphics[width=0.12\textwidth]{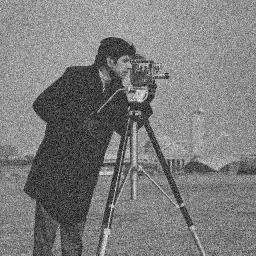}}
  \hspace{-0.3cm}\setlength{\fboxrule}{0.001cm}
  \setlength{\fboxsep}{0cm}
  \fbox{\includegraphics[width=0.12\textwidth]{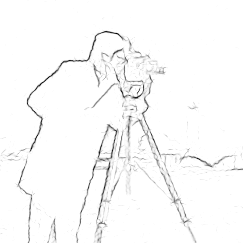}}
  \hspace{-0.3cm}\setlength{\fboxrule}{0.001cm}
  \setlength{\fboxsep}{0cm}
  \fbox{\includegraphics[width=0.12\textwidth]{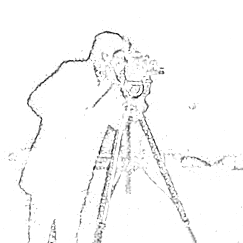}}
  \hspace{-0.3cm}\setlength{\fboxrule}{0.001cm}
  \setlength{\fboxsep}{0cm}
  \fbox{\includegraphics[width=0.12\textwidth]{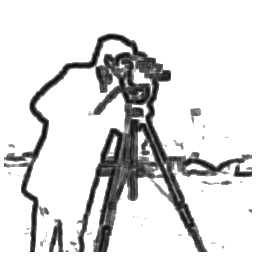}}\\[1pt]
  \hspace{-0.3cm}\setlength{\fboxrule}{0.001cm}
  \setlength{\fboxsep}{0cm}
  \fbox{\includegraphics[width=0.12\textwidth]{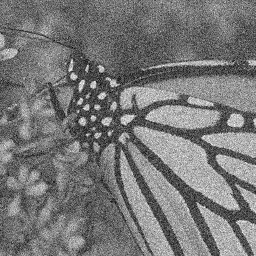}}
  \hspace{-0.3cm}\setlength{\fboxrule}{0.001cm}
  \setlength{\fboxsep}{0cm}
  \fbox{\includegraphics[width=0.12\textwidth]{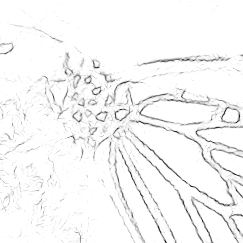}}
  \hspace{-0.3cm}\setlength{\fboxrule}{0.001cm}
  \setlength{\fboxsep}{0cm}
  \fbox{\includegraphics[width=0.12\textwidth]{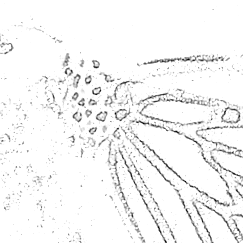}}
  \hspace{-0.3cm}\setlength{\fboxrule}{0.001cm}
  \setlength{\fboxsep}{0cm}
  \fbox{\includegraphics[width=0.12\textwidth]{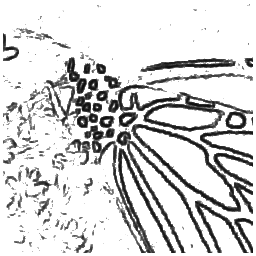}}
  \hspace{-0.3cm}\setlength{\fboxrule}{0.001cm}
  \setlength{\fboxsep}{0cm}
  \fbox{\includegraphics[width=0.12\textwidth]{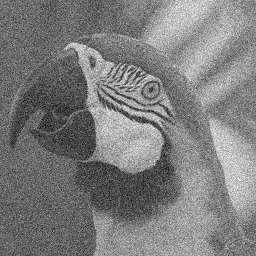}}
  \hspace{-0.3cm}\setlength{\fboxrule}{0.001cm}
  \setlength{\fboxsep}{0cm}
  \fbox{\includegraphics[width=0.12\textwidth]{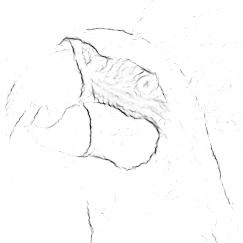}}
  \hspace{-0.3cm}\setlength{\fboxrule}{0.001cm}
  \setlength{\fboxsep}{0cm}
  \fbox{\includegraphics[width=0.12\textwidth]{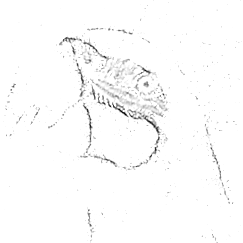}}
  \hspace{-0.3cm}\setlength{\fboxrule}{0.001cm}
  \setlength{\fboxsep}{0cm}
  \fbox{\includegraphics[width=0.12\textwidth]{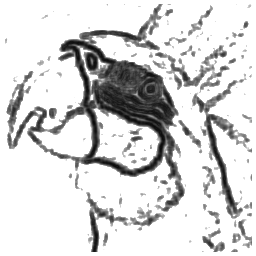}}\\[1pt]
  \hspace{-0.3cm}\setlength{\fboxrule}{0.001cm}
  \setlength{\fboxsep}{0cm}
  \fbox{\includegraphics[width=0.12\textwidth]{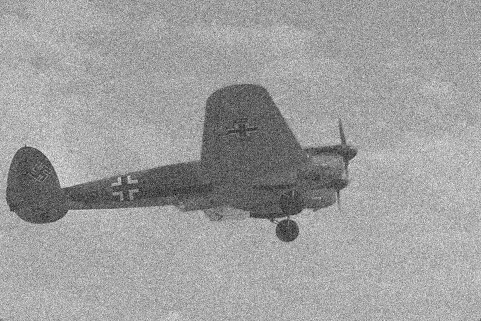}}
  \hspace{-0.3cm}\setlength{\fboxrule}{0.001cm}
  \setlength{\fboxsep}{0cm}
  \fbox{\includegraphics[width=0.12\textwidth]{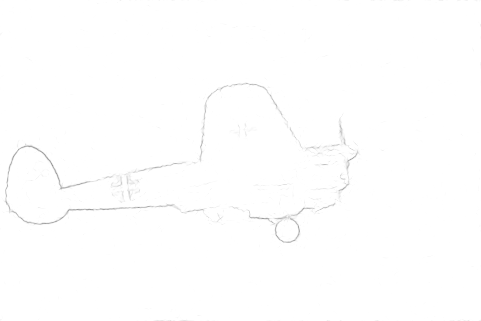}}
  \hspace{-0.3cm}\setlength{\fboxrule}{0.001cm}
  \setlength{\fboxsep}{0cm}
  \fbox{\includegraphics[width=0.12\textwidth]{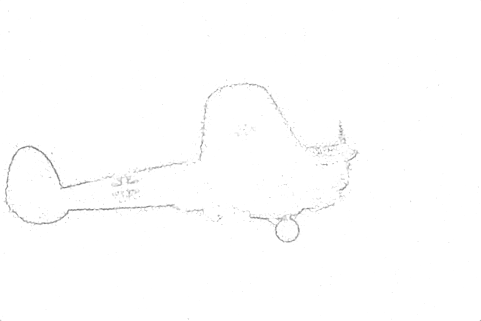}}
  \hspace{-0.3cm}\setlength{\fboxrule}{0.001cm}
  \setlength{\fboxsep}{0cm}
  \fbox{\includegraphics[width=0.12\textwidth]{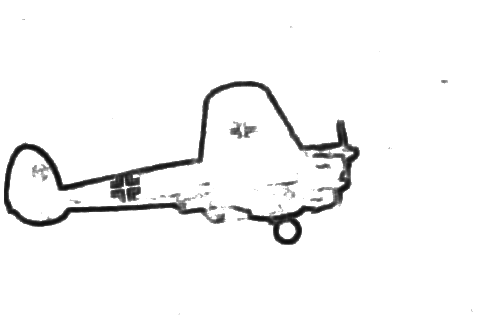}}
  \hspace{-0.3cm}\setlength{\fboxrule}{0.001cm}
  \setlength{\fboxsep}{0cm}
  \fbox{\includegraphics[width=0.12\textwidth]{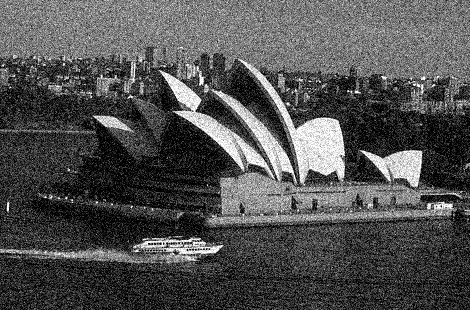}}
  \hspace{-0.3cm}\setlength{\fboxrule}{0.001cm}
  \setlength{\fboxsep}{0cm}
  \fbox{\includegraphics[width=0.12\textwidth]{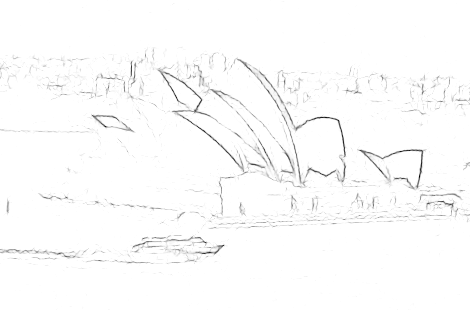}}
  \hspace{-0.3cm}\setlength{\fboxrule}{0.001cm}
  \setlength{\fboxsep}{0cm}
  \fbox{\includegraphics[width=0.12\textwidth]{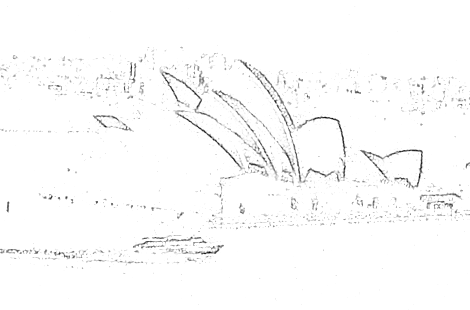}}
  \hspace{-0.3cm}\setlength{\fboxrule}{0.001cm}
  \setlength{\fboxsep}{0cm}
  \fbox{\includegraphics[width=0.12\textwidth]{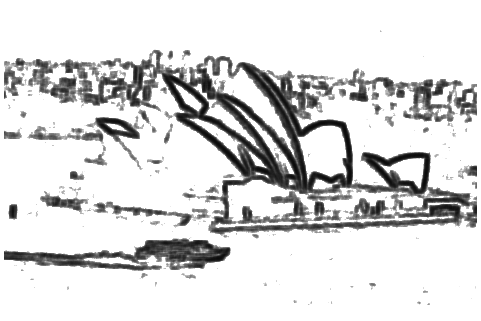}}\\[1pt]
  \caption{from left to right, every four columns are the noisy images, with the noise variance $\sigma = 30$, and the edge strength map based on Log-Gabor transform, monogenic signal and the proposed method respectively.}
  \label{fig:30}
\end{figure*}

The spectrum congruency is obtained according to the local energy model, which is analogous to phase congruency, hence, we compare our results to the Log-Gabor-based and monogenic signal-based phase congruency on some real images.
In the noisy cases, three noise variances are added to the original images, $\sigma = 10, 20, 30$ respectively.
The patch size is $S_1=3\times3$, $S_2=5\times5$, $S_3=7\times7$ for the noise-free image and $\sigma = 10$;
When $\sigma = 20$ and $30$, we choose $S_1=5\times5$, $S_2=7\times7$, $S_3=9\times9$.
The experiment results are displayed in \f\ref{fig:ori}, \f\ref{fig:10}, \f\ref{fig:20} and \f\ref{fig:30}.

Figure.\ref{fig:ori} shows the performance on the noise-free images.
From the results we can see that our method provides much more useful details compared to the other two methods.
For example, in the \emph{opera house} image, the lines and corners of the roof are well detected by our method, while discontinuous and inconspicuous edges are generated by the other two methods.
In addition, even those weak edges can be detected by using our method, such as the background of the parrot image.
Moreover, with the thinning step by using the non-maximum suppression, our method provides more precise location of the edges.

Figure.\ref{fig:10}, \f\ref{fig:20} and \f\ref{fig:30} demonstrate the effectiveness of the spectrum congruency on noisy images, of which the standard deviation equals to $10$, $20$ and $30$ respectively.
As the noise increases, the phase congruency value decreases dramatically by using the Log-Gabor-based method.
When the noise variance adds up to $30$, the feature is nearly unseen.

For the \emph{house} image in \f\ref{fig:10}, a lot of features disappear by using the Log-Gabor filter, the glitch artifact is generated by using monogenic signal, but our method provides stable result as it is in \f\ref{fig:ori}.
The buildings in the \emph{cameraman} image in \f\ref{fig:10} are barely seen by using the Log-Gabor and monogenic signal, while in our case, they are still well detected.
The complete contours of the \emph{parrot}, the \emph{plane} and the\emph{ opera house} are presented by our method, while by using the other two methods, a lot of segments are missing.

With a stronger noise $\sigma=20$ in \f\ref{fig:20}, we can see that our method still yields strong edge strength, while the detected features by the other two methods are weak.
Even though that our result is a little affected by the noise, it still provides as much essential and complete features as possible.
For example, in the \emph{butterfly} image, the other two methods are not able to detect the flowers in the background, while our approach still succeeds in providing these details.
Additionally, it can be seen clearly that with the increase of noise, both phase congruency generated by the Log-Gabor and the monogenic signal decreases a lot.
Even in the places where there exist strong edges, the value of phase congruency is still small.
While in our case, the value of features keeps stable regardless of noise.

When a higher level noise $\sigma=30$ is added, \f\ref{fig:30} demonstrates that the proposed method still achieves good performance.
In the \emph{plane} image, the contour of plane is completely detected by using our proposed method (d), while in (b) and (c), nearly nothing can be observed.
In the\emph{ opera house} image, the buildings in the background is detected and preserved by using our method.

The experimental results show that our proposed spectrum congruency is superior to the phase congruency in detecting features either in noise-free cases or in noisy instances.

\subsubsection{Edge-Thinning Maps}
The edge-thinning maps are always the output of logic operations, and they stand for the exact location of detected edges.
Hence, the FOM measure is used here to evaluate the precision of the location of different edge-thinning maps.
The test is implemented on a synthetic image because synthetic images are equipped with the ideal and uncontroversial groundtruth.
We add different noise deviations $\sigma = 30, 50, 100, 150$ to the test image.
Three state-of-the-art methods are used as comparisons: the Canny edge detector (CED) \cite{Canny86}, the scale multiplication Canny edge detector (SMED) \cite{Bao2005Canny} and the detector based on isotropic and anisotropic Gaussian kernel (IAGK) \cite{2012Noise}.
The results of the three comparison methods are obtained by selecting the optimal parameters which result in the highest FOM.
Since the noise is strong in this experiment, we preprocess the image with a Gaussian filter before computing the spectrum congruency.
The window size of the Gaussian kernel is $7\times7$, and the standard deviation is $s = 3.5$.

\begin{figure*}[htp]
  \centering
  \begin{tabular}{cccccc}
  \multicolumn{1}{c}{$\sigma = 30$}&
  \hspace{-0.3cm}\setlength{\fboxrule}{0.001cm}
  \setlength{\fboxsep}{0cm}
  \fbox{\includegraphics[width=0.16\textwidth]{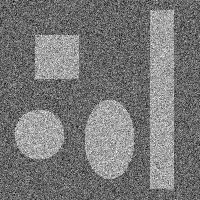}}&
  \hspace{-0.3cm}\setlength{\fboxrule}{0.001cm}
  \setlength{\fboxsep}{0cm}
  \fbox{\includegraphics[width=0.16\textwidth]{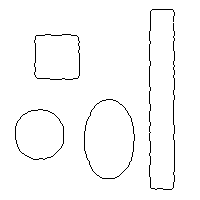}}&  \hspace{-0.3cm}\setlength{\fboxrule}{0.001cm}
  \setlength{\fboxsep}{0cm}
  \fbox{\includegraphics[width=0.16\textwidth]{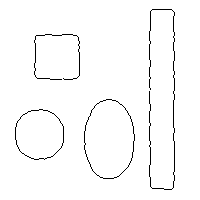}}&
  \hspace{-0.3cm}\setlength{\fboxrule}{0.001cm}
  \setlength{\fboxsep}{0cm}
  \fbox{\includegraphics[width=0.16\textwidth]{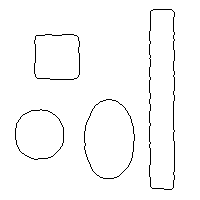}}&
  \hspace{-0.3cm}\setlength{\fboxrule}{0.001cm}
  \setlength{\fboxsep}{0cm}
  \fbox{\includegraphics[width=0.16\textwidth]{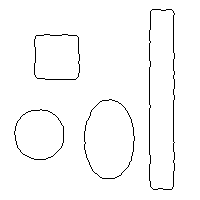}}\\[-1pt]
  $\sigma = 50$&
  \hspace{-0.3cm}\setlength{\fboxrule}{0.001cm}
  \setlength{\fboxsep}{0cm}
  \fbox{\includegraphics[width=0.16\textwidth]{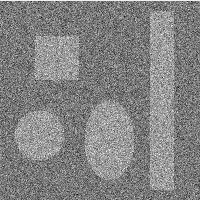}}&
  \hspace{-0.3cm}\setlength{\fboxrule}{0.001cm}
  \setlength{\fboxsep}{0cm}
  \fbox{\includegraphics[width=0.16\textwidth]{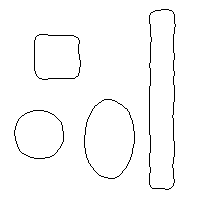}}&
  \hspace{-0.3cm}\setlength{\fboxrule}{0.001cm}
  \setlength{\fboxsep}{0cm}
  \fbox{\includegraphics[width=0.16\textwidth]{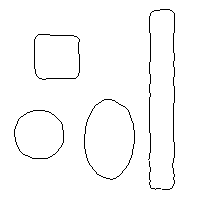}}&
  \hspace{-0.3cm}\setlength{\fboxrule}{0.001cm}
  \setlength{\fboxsep}{0cm}
  \fbox{\includegraphics[width=0.16\textwidth]{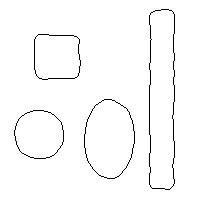}}&
  \hspace{-0.3cm}\setlength{\fboxrule}{0.001cm}
  \setlength{\fboxsep}{0cm}
  \fbox{\includegraphics[width=0.16\textwidth]{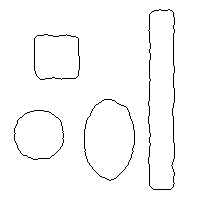}}\\[-1pt]
  $\sigma = 100$&
  \hspace{-0.3cm}\setlength{\fboxrule}{0.001cm}
  \setlength{\fboxsep}{0cm}
  \fbox{\includegraphics[width=0.16\textwidth]{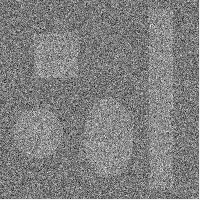}}&
  \hspace{-0.3cm}\setlength{\fboxrule}{0.001cm}
  \setlength{\fboxsep}{0cm}
  \fbox{\includegraphics[width=0.16\textwidth]{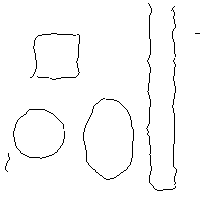}}&
  \hspace{-0.3cm}\setlength{\fboxrule}{0.001cm}
  \setlength{\fboxsep}{0cm}
  \fbox{\includegraphics[width=0.16\textwidth]{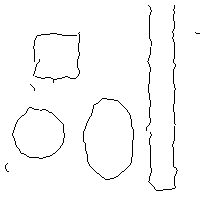}}&
  \hspace{-0.3cm}\setlength{\fboxrule}{0.001cm}
  \setlength{\fboxsep}{0cm}
  \fbox{\includegraphics[width=0.16\textwidth]{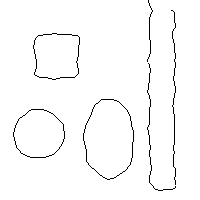}}&
  \hspace{-0.3cm}\setlength{\fboxrule}{0.001cm}
  \setlength{\fboxsep}{0cm}
  \fbox{\includegraphics[width=0.16\textwidth]{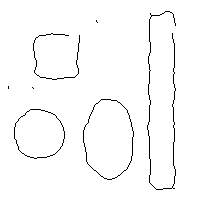}}\\[-1pt]
  $\sigma = 150$&
  \hspace{-0.3cm}\setlength{\fboxrule}{0.001cm}
  \setlength{\fboxsep}{0cm}
  \fbox{\includegraphics[width=0.16\textwidth]{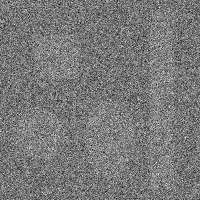}}&
  \hspace{-0.3cm}\setlength{\fboxrule}{0.001cm}
  \setlength{\fboxsep}{0cm}
  \fbox{\includegraphics[width=0.16\textwidth]{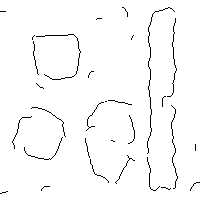}}&
  \hspace{-0.3cm}\setlength{\fboxrule}{0.001cm}
  \setlength{\fboxsep}{0cm}
  \fbox{\includegraphics[width=0.16\textwidth]{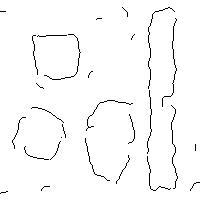}}& \hspace{-0.3cm}\setlength{\fboxrule}{0.001cm}
  \setlength{\fboxsep}{0cm}
  \fbox{\includegraphics[width=0.16\textwidth]{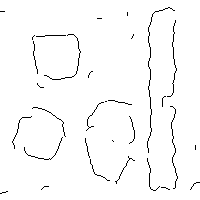}}&
  \hspace{-0.3cm}\setlength{\fboxrule}{0.001cm}
  \setlength{\fboxsep}{0cm}
  \fbox{\includegraphics[width=0.16\textwidth]{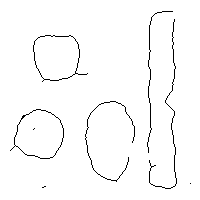}}\\[-1pt]
  &(a)&(b)&(c)&(d)&(e)
  \end{tabular}
  \vspace{-0.2cm}
  \caption{from left to right, top to bottom are: (a) the noisy images, with the noise variance $\sigma = 30, 50, 100, 150$, (b)-(e) the corresponding edge thinning maps obtained by the CED, SMED, IAGK and the proposed method respectively.}
  \label{fig:noisycase}
\end{figure*}

Figure \ref{fig:noisycase} and Table.\ref{tab:performance_comparison} show the experimental results and the FOM index respectively.
Table.\ref{tab:performance_comparison} demonstrates that the proposed method is competitive with the state-of-the-art edge detection method, especially when the noise deviation is high.
In \f\ref{fig:noisycase}, when the noise deviation is low, all the methods can provide precise locations and complete edges.
When there is a high level of noise, the differential-based methods need large scales to confront the noise, while large scales will blur the details, thus leading to some missed detection.
For example, when $\sigma =100$, the top line of the rectangle is detected by our method, while it is missed by using the other three differential-based method.
When $\sigma =150$, compared with the differential-based methods, our method is less affected by the noise and can still provide relatively complete contour of the shapes in the image, while the other methods detect some false segments and incomplete contours.
The experiment results demonstrate that the spectrum congruency is highly robust to noise.

\begin{table}[htb]
\caption{FOM of CED, SMED, IAGK and the proposed method under different noise standard deviation in \f\ref{fig:noisycase}}
\label{tab:performance_comparison}
\centering
\begin{tabular}{ccccc}
\hline
             & CED              & SMED             & IAGK       & Proposed        \\ \hline
$\sigma=30$  &  \textbf{ 0.8195}&  0.8149          & 0.8141     &  0.8137         \\ \hline
$\sigma=50$  &   0.771          &  \textbf{0.7863 }& 0.7767     &  0.7778         \\ \hline
$\sigma=100$ &   0.6978         & 0.7005           & 0.6782     &  \textbf{0.7081}\\ \hline
$\sigma=150$ &   0.6039         & 0.5995           & 0.5991     &  \textbf{0.6138}\\ \hline
\end{tabular}
\end{table}

\subsection{Experiment on BSDS500 Dataset}
\label{sec:comparison}
\begin{figure}
\centering
  {\includegraphics[width=0.45\textwidth]{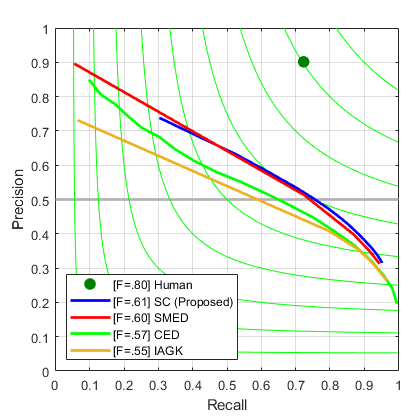}}\\
  \caption{PR curves of the four methods based on the BSDS500 dataset}
  \label{fig:chart}
\end{figure}

The BSDS500 dataset is a widely used dataset for image contour detection and segmentation.
Each image is labeled by $4$ to $9$ annotators manually.
It contains 200 training images, 200 test images, 100 validation images.
Here we use the 100 gray validation images for experiment.
Each method is configured as follows:

\begin{itemize}
  \item CED: The scale is $\sqrt{2}$, which is widely adopted in the literature.
  \item SMED: The two scales are $2$ and $8$ respectively.
  \item IAGK: The scale is specified as $4$, the anisotropic factor is $2\sqrt{2}$ and the number of kernel orientation is 16.
  \item SC: The patchsize is $3, 5, 7$ respectively, $\alpha=0.05$.
\end{itemize}
The PR vurves of the four edge detectors are displayed in \f\ref{fig:chart}.
In terms of the PR curve, we can see that our proposed edge detector behaves better than the CED, SMED and IAGK approaches.

Meanwhile, Table.\ref{tab:performance_bsds500} indicates that the proposed method achieves better performance.
The $F_{ODS}$, $F_{OIS}$ and $R50$ indexes demonstrate that our proposed method is competitive with the other three methods.

\begin{table}[htb]
\caption{F-measure of CED, SMED, IAGK and the proposed method}
\label{tab:performance_bsds500}
\centering
\begin{tabular}{ccccc}
\hline
           & CED     &  IAGK    & SMED    & Proposed         \\ \hline
$F_{ODS}$  & 0.566   &  0.549   & 0.601   & \textbf{0.607}   \\ \hline
$F_{OIS}$  & 0.608   &  0.543   & 0.610   & \textbf{0.626}   \\ \hline
$R50$      & 0.649   &  0.589   & 0.741   & \textbf{0.756}   \\ \hline
\end{tabular}
\end{table}

\section{Conclusions and Future Work}
\label{sec:conclusion}
This paper proposes a novel image feature called spectrum congruency for edge detection.
The spectrum congruency is computed based on the local energy of multiscale patches.
Similar to the phase congruency, spectrum congruency is also consistent with the human visual system on perceiving the features of interest in the images.
Unlike phase congruency, which measures the local energy via the transforms of fixed bases such as the Fourier transform, Log-Gabor transform and the monogenic signals, the proposed spectrum congruency computes the local energy in a data-driven way, which are more adaptable to the input images.
Patches with different sizes around each pixel are selected and re-sampled to acquire information from low to high frequency information.
The spectrum congruency indicates the edge strength map and is computed by integrating information at different frequency band.

Compared with the previous measurement of phase congruency, our method is more robust to noise, with fewer glitch artifacts, smoother and more continuous edge features.
Besides, compared with state-of-the-art edge detectors, our method can provide more reliable edges when confronting high level of noise.
In future, we are going to apply our edge detection method higher level computer vision tasks such as image registration, image segmentation, object recognition \emph{etc.}

\bibliographystyle{ieeetr}
\bibliography{egbib}

\end{document}